\documentclass[journal]{IEEEtran}
\usepackage{url}
\usepackage{times}
\usepackage{bm}
\usepackage{theorem}
\usepackage{amsmath}
\usepackage{latexsym}
\usepackage{amssymb}
\usepackage{graphics}
\usepackage{graphicx}

\makeatletter
\def\eqnarray{\stepcounter{equation}\let\@currentlabel=\theequation
\global\@eqnswtrue
\global\@eqcnt\z@\tabskip\@centering\let\\=\@eqncr
$$\halign to \displaywidth\bgroup\@eqnsel\hskip\@centering
  $\displaystyle\tabskip\z@{##}$&\global\@eqcnt\@ne 
  \hfil$\;{##}\;$\hfil
  &\global\@eqcnt\tw@ $\displaystyle\tabskip\z@{##}$\hfil 
   \tabskip\@centering&\llap{##}\tabskip\z@\cr}
\makeatother
\theorembodyfont{\itshape}
\newtheorem{lem}{Lemma}[section]
\newtheorem{prop}{Proposition}[section]
\theorembodyfont{\rmfamily}
{\bf}{\rm}
{\bf}{\rm}
{\bf}{\rm}

\newtheorem{rem}{Remark}[section]{\itshape}{\rmfamily}

\newcommand{\qed}{\hspace*{\fill}$\Box$}

\newcommand{\rmd}{\mathrm{d}}
\newcommand{\calL}{\mathcal{L}}

\newcommand{\bbN}{\mathbb{N}}
\newcommand{\bbR}{\mathbb{R}}

\newcommand{\bbE}{\mathbb{E}}

\newcommand{\cosec}{\mathrm{cosec}}

\newcommand{\PP}{\mathsf{P}}

\newcommand{\dd}[1]{\if#11 1\!\!1 
\else {\if#1C I\!\!\!C
\else {\if#1G I\!\!\!G 
\else {\if#1J J\!\!\!J 
\else {\if#1S S\!\!\!S
\else {\if#1Z Z\!\!\!Z
\else {\if#1Q O\!\!\!\!Q
\else I\!\!#1
\fi} 
\fi}
\fi}
\fi} 
\fi} 
\fi} 
\fi} 
\begin{document}
\title{Theoretical Performance Analysis of Vehicular Broadcast Communications at Intersection
and their Optimization}

\author{Tatsuaki~Kimura~%\IEEEmembership{Member,~IEEE,}
        and~Hiroshi~Saito~%\IEEEmembership{Fellow,~IEEE}
%        and~Jane~Doe,~\IEEEmembership{Life~Fellow,~IEEE}% <-this % stops a space
\thanks{T.~Kimura is with 
Department of Information and Communications Technology, 
Graduate School of Engineering, Osaka University, 
Osaka, Japan (kimura@comm.eng.osaka-u.ac.jp), and
 H.~Saito is with Mathematics and Informatics Center, University of Tokyo, Tokyo, Japan. }%
%        Network Technology Laboratories, NTT Corporation, Tokyo 180--8585, Japan
%        (e-mail: kimura.tatsuaki@lab.ntt.co.jp)}% <-this % stops a space
%\thanks{Manuscript received XX XX, 20XX; revised XX XX, 20XX.}
%\thanks{This  work  has  been  submitted  to  the  IEEE  for  possible  
%publication. }
%\thanks{
%\copyright~2017 IEEE. Personal use of this material is permitted. Permission from IEEE must be obtained for all other uses, in any current or future media, including reprinting/republishing this material for advertising or promotional purposes, creating new collective works, for resale or redistribution to servers or lists, or reuse of any copyrighted component of this work in other works.
%}
}

\maketitle

\begin{abstract}
Cooperative vehicle safety (CVS) systems are a key application of intelligent transportation systems
because they include many applications, such as cooperative collision warning. % and emergency brake lights. 
In CVS systems, vehicles periodically broadcast their information, e.g., position and speed. 
In this paper, we propose an optimization method for the broadcast rate in 
vehicle-to-vehicle (V2V) broadcast communications at an intersection
on the basis of theoretical analysis. 
%theoretically analyze the performance of vehicle-to-vehicle (V2V) broadcast communications at an intersection and provide tractable formulae of performance metrics to optimize them. 
%
We consider a model in which locations of vehicles are modeled separately as {\it queuing} and {\it running} segments and derive key performance metrics of V2V broadcast communications  via a stochastic geometry approach. 
Since these theoretical expressions are mathematically intractable, we developed {\it closed-form} approximate formulae for them. 
%In our approximation, the probability of successful transmission decreases geometrically as the distance to a receiver increases and is expressed by only system parameters. 
Using them, we optimize the broadcast rate such that the mean number of successful receivers per unit time is maximized. 
Because of the closed form approximation, the optimal rate can be used as a guideline for a {\it real-time} control-method, which is not achieved through time-consuming simulations. 
We evaluated our method through numerical examples and 
demonstrated the effectiveness of our method. 

\end{abstract}

%\begin{IEEEkeywords}
%ITS; V2V communication; Cooperative vehicle safety (CVS); VANET;
% broadcast; stochastic geometry; 
%Poisson point process; broadcast rate control; intersection
%\end{IEEEkeywords}
%

% \fi
%\IEEEpeerreviewmaketitle

\section{Introduction}

%\IEEEPARstart{I}{ntelligent}
Intelligent transportation systems (ITSs) are promising technology for improving safety for drivers/pedestrians and the efficiency of transportation~\cite{Hart09}. In general, vehicle-to-infrastructure (V2I) and vehicle-to-vehicle (V2V) communications play a key role in achieving ITSs. These communications are commonly based on narrow-band dedicated short range protocols (DSRC). For instance, wireless access in vehicular environments (WAVE) is the protocol suite adopted in the U.S. In WAVE, IEEE 802.11p~\cite{80211p06} is standardized for the media access control (MAC) and physical layers. %similar to IEEE~802.11a. 
% and uses 5.9-GHz bandwidth and carrier sense multiple access (CSMA) similar to IEEE~802.11a. %In addition, IEEE~1609~\cite{160907} designs higher-layer functions such as networking and multichannel operations. 

Cooperative vehicle safety (CVS) systems~\cite{CVS} are one of the key applications of ITSs
using V2V communications. CVS systems include many applications such as cooperative collision warning and emergency brake lights~\cite{Huan10}. In these systems, vehicles periodically broadcast their information e.g., positions (Global Positioning System; GPS), speed, and braking status, so that vehicles can track the positions of other vehicles and avoid traffic congestion, collisions, or unknown hazards. CVS systems have been attracting much attention in recent decades because these applications  will drastically change our lives. 

Because of the critical nature of CVS systems, their performance analysis and management are hot research topics. Broadcasting with a high transmission power and high broadcast rate in congested roadways may significantly degrade the wireless communication quality due to high interference. To reduce the interference caused by a large number of vehicles sharing the same channel, several schemes have recently been proposed to adaptively control the transmission power or broadcasting rate~\cite{Huan10, Fall10, Fall16, Tiel11, Torr09, Mitt08}. 
However, most of these schemes are not based on theoretical analysis and are commonly evaluated through simulations. 
Because the environments in which V2V communications occur may quickly and frequently change, a more general understanding of performance is crucial to effectively control CVS systems. 
Furthermore, most studies consider only {\it homogeneous} environments, such as multi-lane highways, in which vehicles are distributed with the same traffic density. However, to deploy CVS systems in urban environments, more realistic {\it inhomogeneous} situations, such as intersections, must be taken into account. More specifically, the density of vehicles near an intersection is much higher than that on a normal road due to queuing vehicles and crossing streets, and thus the interference near the intersection also becomes much higher. 
As a result, the communication quality at an intersection is very different from that in homogeneous environments. 
Recently, an optimization of transmission power of vehicles at an intersection was theoretically analyzed~\cite{Kimu18}. However, the obtained analytical results are highly complicated and mathematically intractable, and thus the analysis cannot be applied to real-time control due to its high computational time. 

In this paper, we propose an optimization method for V2V broadcast communications at an intersection
on the basis of theoretical analysis. %of V2V broadcast communications at an intersection and derive performance metrics in a tractable form. 
By deriving performance metrics of V2V broadcast communications and 
expressing them as tractable approximate formulae, we can optimize the broadcast rate in a reasonable computational time so that the number of successful receivers per unit time is maximized. 
We consider an intersection model, in which locations of vehicles are separated into {\it queuing} segments and {\it running} segments. In the former, vehicles are assumed to be queuing at even intervals; and in the latter, vehicles are distributed in accordance with a homogeneous Poisson point process (PPP). By using a stochastic geometry approach, theoretical values are derived for the two key performance metrics of V2V broadcast communications: the probability of successful transmission and the mean number of successful receivers. 
The former is defined as the probability that the signal-to-interference-ratio (SIR) of a receiver exceeds a certain threshold, and the latter as the expected number of vehicles that can successfully receive information from a transmitter. 
However, these results from exact analysis are expressed in {\it non-analytical} form and require time-consuming numerical computation. Thus, they are too complicated for not only the forms of the function to be understood but also their system parameters to be optimized. To address this problem, we developed a closed-form approximation for the performance metrics %of V2V broadcast communications. 
by assuming sufficiently large queues. 
%More precisely, when vehicles broadcast to their neighbors, the probability of successful transmission geometrically decreases as the distance to the receiver increases in our approximation. 
%In addition, the decay rate depends only on a few system parameters, such as the intensity of vehicles. 
Using the approximate formulae, we optimize the broadcast rate of vehicles that maximizes the number of successful receivers per unit time. 
The closed-form expression enables us to easily compute  the optimal broadcast rate without time-consuming numerical computation. 
Therefore, our optimization method can be applied to {\it real-time} broadcast rate control for CVS systems to mitigate the interference problem caused by congestion.
Numerical results revealed the proposed optimization could mitigate the interference problem at an intersection. 
We also found that our approximation fitted well to both simulation and exact analysis.

%We briefly summarize our contributions as follows.
%%
%\begin{itemize}
%%
%\item We developed a closed-form approximation for the theoretical values of 
%two key performance metrics
%%theoretically analyzed 
%of the performance of V2V broadcast communications at an intersection: 
%the probability of successful transmission and the mean number of successful receivers.
%The approximate formulae depend only on system parameters and thus do not 
%require time-consuming numerical computation. 
%
%%
%\item By using the approximation formulae, we developed a method 
%for optimizing the broadcast rate of vehicles. 
%Balancing the broadcast rate and the mean number of successful receivers, 
%we improved V2V broadcast communications at a congested intersection. 
%In accordance with the closed-form formulae, the optimal broadcast rate can be 
%obtained in a reasonable computational time and thus can be applied
%to real-time  broadcast rate control. 
%%
%\item We evaluated our approximation and optimization methods
%through simulation and numerical experiments. 
%We found that our optimization method had sufficient performance. 
%We also found that our approximation fitted
%well to the results from simulation and exact analysis in a realistic setting. 
% 
%%
%\end{itemize}
%

The remainder of this paper is organized as follows. 
Section~\ref{sec-related} summarizes previous studies. 
In Section~\ref{sec-model}, we explain the system model considered in this paper. 
%We provide an exact analysis in Section~\ref{sec-anal-exact}. 
Section~\ref{sec-app} presents the approximate analysis 
of the key performance metrics of V2V communications at an intersection.
In Section~\ref{subsec-optimize}, we provide a broadcast-rate-optimization method based on the analytical results.
Finally, we discuss several numerical experiments in Section~\ref{sec-experiments}, 
and conclude the paper in Section~\ref{sec-conclude}.

\section{Related work}\label{sec-related}
Due to the importance of ITSs, there have been a lot of studies 
in the area of the performance evaluation of V2I/V2V communications in the past 
decade. Most of the earlier work is simulation-based~\cite{Torr09, Torr04, Eich07, Burg08}. 
However, simulation-based approaches often require much computational time and resources. 
The previous work~\cite{Fall11, Gian12, Han12, Yao13} conducted a theoretical analysis of 
the CSMA behaviors of IEEE 802.11p on the basis of a Markov chain model approach. 
Fallah et al.~\cite{Fall11} studied the impact of the rate and range of broadcasting 
on network performance in a highway environment considering the hidden terminal problem. 
Han et al.~\cite{Han12} and Yao et al.~\cite{Yao13} analyzed the enhanced distributed channel 
access (EDCA) behavior in IEEE 802.11p, in which different access categories have different 
contention windows and arbitration inter-frame space. 
However, these studies did not consider the geographical effects or interference 
in V2V communications and assumed only simple communication scenarios. 

To reduce the interference of V2V broadcast communications, 
several adaptive control schemes for 
transmission power \cite{Huan10, Torr09, Mitt08} or
broadcasting rate~\cite{Huan10, Fall10, Fall16, Tiel11} have recently been proposed. 
The method proposed by Moreno et al.~\cite{Torr09} adaptively controls 
the transmission power of vehicles so that their max-min fairness is satisfied.
In \cite{Mitt08}, a segment-based power control method based on a distributed 
vehicle density estimation algorithm is proposed. 
Huang et al.~\cite{Huan10} developed broadcast rate and power control algorithms,
in which the rate is determined by estimating the channel error rate and
the power is determined by observing the channel status. 
Tielert et al.~\cite{Tiel11} introduced a rate adaptation algorithm based on the channel
busy ratio. Most recently, Fallah et al.~\cite{Fall16} 
updated the algorithm of \cite{Fall10}
so that the power changes in each iteration can be configurable and stable. 
None of the adaptive control methods above was based on theoretical interference analysis
and were considered in simple environments such as multi-lane highways, 
in which vehicles are running in the same direction with the same
traffic density.
However, theoretical guidelines for more realistic situations, such as intersections,
are crucial to deploy CVS systems in more complex urban environments.

Stochastic geometry is a powerful mathematical tool for modeling random spatial events and has been applied to the area of %wireless communication~\cite{Haen13} including 
vehicular networks~\cite{Blas13, Faro15, Nguy13, Tong16, Stei15, Kimu18, Chet18, Choi18, Nguy07}. 
By modeling the locations of communication devices, such as vehicles and road side units (RSUs), as a spatial point process, theoretical values of various performance metrics can be calculated. Such mathematical understanding of the ITS system not only frees us from time consuming simulation but also helps in optimizing system parameters or analyzing their sensitivity. 
In previous studies~\cite{Nguy13, Tong16}, the behavior of CSMA used in DSRC was analyzed. 
More specifically, Nguyen et~al.~\cite{Nguy13} showed that CSMA behaves like an ALOHA-type transmission pattern in dense networks and derived the theoretical expression of performance metrics in broadcast V2V communications while assuming that vehicles are distributed in accordance with spatially homogeneous PPP. 
In addition, Tong et al. \cite{Tong16} studied the performance of DSRC in both the spatial and time domains by using a Markov chain model approach for CSMA, which is similar to that of Nguyen et~al.~\cite{Nguy07}. 
More recently, Chetlur and Dhillon~\cite{Chet18} studied V2V communications where 
vehicles are distributed on  roads that is randomly distributed according to Poisson line process. 
Similarly, by considering the spatial patterns of and vehicles on roads and  cellular base stations together, 
Choi and Baccelli~\cite{Choi18} analyzed the coverage probability of cellular-assisted vehicular communications. 
%However, these studies assumed homogeneously distributed vehicles and did not consider approximation and
%power or broadcast rate control. 
%
However, the above studies considered only homogeneous situations and 
did not consider power or broadcast rate control. 
Similar to us, Steinmetz et al.~\cite{Stei15} analyzed packet reception probability at an intersection by modeling the locations of vehicles as a homogeneous PPP. They also considered an inhomogeneous PPP scenario as an extension, but 
no specific intensity function of vehicular density was given. 
In our previous study~\cite{Kimu18}, we directly modeled  the queueing segment in an intersection and 
proposed optimization of transmission power based on theoretical analysis. 
However, the obtained analytical results are highly complicated and mathematically intractable. 
Contrary  to these studies, we propose a real-time broadcast rate optimization method
by deriving tractable results. 
% but the obtained result is insufficient because no specific intensity function of vehicular density was given and no tractable results were derived that clearly represent the impacts of queues at an intersection. 

\section{Model description}\label{sec-model}
\begin{figure}[!t]
\centering
\includegraphics[width=3.0in]{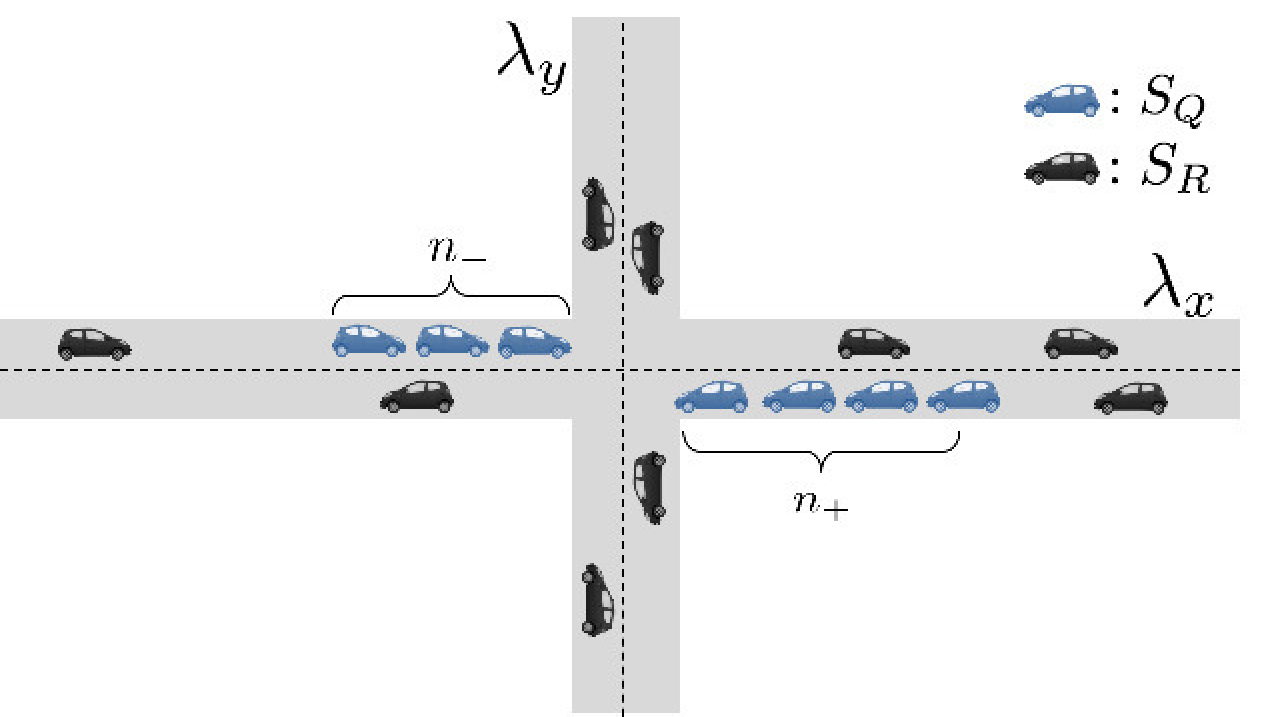}
\caption{System model. Vehicles in running segment ($S_R$)
are distributed in accordance with homogeneous PPP with intensity 
$\lambda_{x}$ or $\lambda_{y}$. 
Intervals of vehicles in queueing segment ($S_Q$) are fixed value $l_v$.
$n_{+}$ and $n_{-}$ represent number of vehicles stopping at intersection. 
}
\label{fig:model}
\end{figure}

In this section, we explain the system model. 
Figure~\ref{fig:model} shows a conceptual image of our model. We consider an intersection where two streets are crossing. One street runs parallel along the $x$-axis, and the other along the $y$-axis. On the street along the $x$-axis, vehicles are queuing, i.e., stopped, at the intersection, and on both streets, vehicles are running. We call these parts a queuing segment $S_Q$ or a running segment $S_R$. In addition, $S_{R_x}$ and $S_{R_y}$ denote the running segments on the $x$- or $y$-axis, respectively. %i.e., $S_R = S_{R_x}\cup S_{R_y}$. 
We assume that vehicles in $S_R$ are distributed in accordance with a homogeneous PPP on each street. Let $\lambda_{x}$ and $\lambda_{y}$ denote the intensity of vehicles in $S_{R_x}$ and $S_{R_y}$. Let $n_{+}$ and $n_{-}$ denote the numbers of vehicles stopped at an intersection in each part, where the subscript $\{+, -\}$ represents the positive or negative part on the $x$-axis. We assume that vehicles have length $l_v$ and the widths of the streets (i.e., those of vehicles) are negligible. Note that there is no queue on the $y$-axis because we consider the case where the traffic signals on the $y$-axis are green. We can %easily consider 
%the case where those on the $x$-axis are green. 
apply the same discussion in this paper to the case where those on the $x$-axis are green. 

We next explain the channel model. Vehicles periodically broadcast a packet 
and each transmission requires $L$ [sec.]. 
%such as GPS, their braking information, or emergency information to neighbors, i.e., V2V communications. 
%All vehicles are assumed to be equipped with devices necessary for V2V communications. 
We assume that vehicles in $S_Q$ independently transmit with rate $\theta\in (0, 1/L)$ [1/sec.] and those in $S_R$ with $\theta_0\in (0, 1/L)$. If time is slotted and each slot size is $L$, then each vehicle transmits at each time-slot in accordance with an independent Bernoulli distribution. More specifically, the probability (i.e., the parameter of Bernoulli distribution) that each vehicle in $S_Q$ (resp. in $S_R$) is transmitting in each time slot is $\rho \triangleq \theta L \in (0, 1)$ [resp. $\rho_0 \triangleq \theta_0 L \in (0, 1)$]. 
Since $\theta$ and $\rho$ ($\theta_0$ and $\rho_0$) have one-to-one correspondence, we only consider 
$\rho$ and $\rho_0$ hereafter.  
We also assume that vehicles currently transmitting cannot receive a packet from other vehicles at the same time. 
The transmission power of all vehicles is normalized to 1. 
Antenna gain is assumed to be equal to 1 throughout this paper. 
In addition, all transmission channels have the effect of Rayleigh fading 
and $h$ denotes the random variable that represents the fading gain. 
The path loss model is $r^{-\alpha}$ for distance $r \in \bbR_+$ and where $\alpha > 1$ is a path loss exponent. Thus, the received power from vehicle $x_i$ at distance $r$ can be expressed as $h_i r^{-\alpha}$. %Table~\ref{tab:logs} summarizes the notations used in this paper. 
Table~\ref{tab:logs} summarizes the notations used in this paper. 

\begin{table}[ht]
\caption{List of notations}
\centering
\begin{tabular}{ll} \hline
$l_v$ & length of vehicle \\
$S_R$ & set of vehicles running on street \\
$S_{R_x}$ & set of vehicles running on street along $x$-axis\\
$S_{R_y}$ & set of vehicles running on street along $y$-axis \\
$S_Q$ & set of vehicles stopping/queueing at  intersection \\
%$L$ & each transmission time \\
%$\theta, \theta_0$ & broadcast rate of vehicles in $S_Q$ and $S_R$ \\
$\rho, \rho_0$ & probability that vehicles in $S_Q$ and $S_R$ are transmitting \\
$\lambda_{z}$ & intensity of vehicles in $S_R$ 
                     $(z \in \{x, y\})$\\
$n_{\ast}$ & number or vehicles stopped at intersection $(\ast \in \{+, -\})$\\               
$h_i$ & fading variable \\
$I_R$ & interference from vehicles in $S_R$s \\
$I_Q$ & interference from vehicles in $S_Q$s \\
 \hline
\end{tabular}
\label{tab:logs}
\end{table}

Note that CSMA is designed as the MAC layer protocol in IEEE 802.11p~\cite{Kara11}. 
Since vehicles that are close to each other do not transmit simultaneously in CSMA, 
hard-core point processes have been used for modeling such CSMA-based 
protocols~\cite{Tong16, Nguy07, Elsa13}; however, they are not mathematically tractable
because they are obtained by {\it dependent} thinning of a PPP. 
%However, models for CSMA networks are not mathematically tractable~\cite{Tong16, Nguy07, Elsa13}. 
In addition, Nguyen et~al.~\cite{Nguy13} claimed that CSMA behaves like 
an ALOHA-type transmission pattern in dense networks. 
This is mainly because there are 
nodes that choose the same back-off counter due to finite collision window size in 
the binary exponential backoff of CSMA~\cite{Nguy13}. 
Indeed, Tong et al.~\cite{Tong16}
showed that results with an ALOHA-type model were similar to those 
obtained by NS2 simulation that models the CSMA behavior in their numerical examples. 
%when $\lambda \in [30, 130]$ in their numerical examples. 
Therefore, 
we assume that transmitting vehicles use 
the ALOHA-type MAC protocol and model the locations of vehicles by a PPP. 
Note that in such a model, each vehicle attempts to transmit a packet 
with a certain probability in each time slot, 
%because those of such transmitting vehicles 
and thus the positions of transmitters 
can be modeled by independent thinning of the original PPP
(see e.g., \cite{Blas13}).

Let $I$ denote a random variable representing 
the total received interference, from all the vehicles. 
If we consider the tagged channel in which the communication 
distance is equal to $r$,  the signal-to-interference-ratio (SIR)
can be written as
$
\mathsf{SIR}_r = h r^{-\alpha} / I.
$
We then define the probability of successful transmission 
as the probability that the SIR
of a tagged receiver exceeds a threshold $T$,
i.e., 
\begin{eqnarray}
p(r) 
&\triangleq&
\PP(\mathsf{SIR}_r > T)
=
\PP\left(
{h r^{-\alpha}  \over I} > T
\right)
\nonumber
\\
&\overset{(a)}{=}&
\bbE_I\left[
\exp\left( -  T r^{\alpha} I\right)
\right]
= \calL_{I}(T r^{\alpha}),\quad
\label{eq-cov}
\end{eqnarray}
where $\calL_{I}(s)$ is the Laplace transform of $I$
and (a) holds due to Rayleigh fading assumption.  
%Probability $p(r)$ is a key performance metric 
%of V2V broadcast communications, and thus, we mainly focus on the
%analysis of this value in our paper. 
%we  the effect of external noise. 

\subsection{Performance Metrics}\label{sec-anal-exact}

%\section{Performance Metrics}\label{sec-anal-exact}
%
In this section, we provide theoretical expressions of 
performance metrics of V2V communications. % via a stochastic geometry approach. 
%and derive the theoretical expression of the probability of successful transmission
%and the mean number of successful receivers. 
%Since equation (\ref{eq-cov}) shows that $p(r)$ can be obtained 
%as the Laplace transform of the interference from vehicles, 
%we first derive them. 
% and then provide 
%the theoretical values of the probability of successful transmission. 

\subsubsection{Interference distributions}
%\subsubsection{Interference from queueing vehicles}\label{subsec-inter-S_Q}
%
We first consider the interference from vehicles in  $S_Q$. 
For this purpose, we assume that a tagged receiver is in the positive part on the $x$-axis 
and at distance $d$ from the intersection.
In addition, let $d_m= |d - m l_v|\ (1 \le m \le n_{-} + n_{+})$ denote the distance between
the tagged receiver and the $m$-th vehicle from the intersection. 
Therefore, the total interference power received from $S_Q$ is 
$
I_Q \triangleq \sum_{m =1}^{n_{-} + n_{+}} h_m \delta_m d_m^{-\alpha},
$
where $\delta_m = 1$ if the $m$-th vehicle transmits, 
and $\delta_m= 0$ otherwise. Recall that $h_m$ is exponential 
with mean $1$ (the Rayleigh fading assumption). 
Recall also that the vehicles in the $S_Q$ are transmitting with the probability $\rho$. 
Therefore, the Laplace transform of $I_Q$, 
$\calL_{I_Q}(s \mid d) \triangleq \bbE_{I_Q}[\exp(- s I_Q) \mid d]$, 
is equal to 
\begin{eqnarray}
\calL_{I_Q}(s \mid d) 
%&\triangleq&
%\bbE[\exp(- I_Q) \mid d] 
%\nonumber
%\\
&=& 
\bbE_{I_Q}\left[\left.\exp\left(- s\sum_{m = 1}^{n_{-} + n_{+}}  h_m \delta_m d^{-\alpha}_m \right)\right|d\right]
\nonumber
\\
&=&
\prod_{m=1}^{n_{-} + n_{+}} \left[ { \rho  \over  1 +  {s \over (d_m)^{\alpha}}}
+ 1 - \rho
\right].
\label{eq-lap-I_Q}
\end{eqnarray}
%
%\subsubsection{Interference from running vehicles}\label{subsec-inter-S_R}
%

We next consider the interference from the vehicles in  $S_R$. 
Similar to the previous case, 
we assume that a tagged receiver is at distance $d$
from the intersection in the positive part on the $x$-axis. 
Let $\Phi^X_R$ and $\Phi_R^Y$ denote PPPs corresponding to  
 $S_{R_x}$ and $S_{R_y}$. 
The total interference from $\Phi^X_R$ can be represented as 
$I_R^X  = \sum_{x_i \in \Phi^X_R} h_i |x_i - d|^{-\alpha}$. 
Recall that vehicles in $S_R$ transmit with probability $\rho_0$. 
By following  a well-known computation of the Laplace functional of 
the Poisson point process
(see e.g., Proposition~1.5 and Corollary~2.9 in~\cite{Bacc09a}), 
we can compute the Laplace transform of $I_R^X$ as follows.
\begin{eqnarray}
%\lefteqn{
\calL_{I^X_R}(s) &\triangleq &
\bbE_{{I^X_R}} \left[ \exp\left({-s \sum_{x_i \in \Phi_R^X }{h_i \delta_i \over |x_i - d|^{\alpha}}}\right) \right]
%} \qquad&&
\nonumber
\\
%&=&
%\exp\left(-\rho_0  \lambda_x \int_{-\infty}^{\infty} 1 - \calL_h\left({s \over \mu x^{\alpha}}\right) \rmd x \right)
%\nonumber
%\\
&=&
\exp\left(- \rho_0 \lambda_x \int_{-\infty}^{\infty}%{\bbR \backslash [-n_{x,-}l_v, n_{x,+}l_v]} 
{s \over  |x|^{\alpha} + s } \rmd x\right)
\nonumber
\\
&=&
\exp\left(- \rho_0 \lambda_x {2 \pi \over \alpha} \cosec\left({\pi  \over \alpha}\right)\right). 
\label{eq-lap^X_I_R}
\end{eqnarray}
%
%where $\calL_h(s)$ in the second equality denotes the Laplace transform of $h$. 
Note that the distance from the tagged transmitter to a vehicle at distance $y$ from the 
intersection on the $y$-axis is equal to $\sqrt{y^2 + d^2}$. 
Thus, if $I_R^Y$ denotes the total interference from $\Phi^Y_R$, we have
$I_R^Y  = \sum_{y_i \in \Phi^Y_R} h_i (y_i^2 + d^2)^{-\alpha/2}$. 
Therefore, similar to (\ref{eq-lap^X_I_R}), we obtain (see also Section~2 in \cite{Kimu18}), 
\begin{eqnarray}
%\lefteqn{
\calL_{I^Y_R}(s \mid d) 
&\triangleq &
\bbE_{I^Y_R} \left[\left. \exp\left({-s \sum_{y_i \in \Phi_R^Y}
{h_i \delta_i \over  (y^2_i + d^2)^{\alpha \over 2}}}\right) \right| d\right]
\nonumber
\\
%&=&
%\exp\left(- \rho_0  \lambda_y \int_{-\infty}^{\infty} 1 - 
%\calL_h \left({s \over \mu (y^2 + d^2)^{\alpha \over 2}}\right) \rmd y\right)
%\nonumber
%\\
&=&
\exp\left(- \rho_0 \lambda_y \int_{-\infty}^{\infty} 
{s \over  (y^2 + d^2)^{\alpha \over 2} + s} \rmd y\right).\quad
\label{eq-lap^Y_I_R}
\end{eqnarray}
\subsubsection{Probability of successful transmission}
Note that  the total interference from all the vehicles
can be represented as  $I = I_Q + I_R^X + I_R^Y$. 
%its Laplace transform is equal to 
%%Note that the Laplace transform of the total interference 
%%from all the vehicles can be represented as
%%
%\begin{eqnarray}
%\calL_I(s) = \bbE[ \exp(- s I )] 
%%&=& \bbE[ \exp(- s(I_Q + I_R^X + I_R^Y))]
%%\nonumber
%%\\
%&=& \calL_{I_Q}(s \mid d)\calL_{I_R^X}(s)\calL_{I_R^Y}(s \mid d).
%\nonumber
%\end{eqnarray}
%%
%where $\calL_{I_Q}(s \mid d)$, $\calL_{I_R^X}(s)$, and $\calL_{I_R^Y}(s \mid d)$
%are given in (\ref{eq-lap-I_Q})--(\ref{eq-lap^Y_I_R}), respectively. 
Thus, by applying this to (\ref{eq-cov}),  
we can easily obtain the probability of successful transmission as follows. 
\begin{prop}\label{prop-cov}
If a transmitter is at distance $d$ from an intersection, 
the probability of successful transmission to a receiver at distance $r$ 
from the transmitter on the $x$-axis is given by
\begin{eqnarray}
p(r) &=& \calL_{I_Q}(T r^{\alpha} \mid d') 
\calL_{I^X_R}(T r^{\alpha} ) \calL_{I^Y_R}(T r^{\alpha}  \mid d'),
\label{eq-cov-full}
\end{eqnarray}
where $d' = d + r$ if the receiver is on the right-hand side of the transmitter,
and $d' = |d - r|$ otherwise. 
\end{prop}

Although Proposition~\ref{prop-cov} only shows 
the case where a transmitter and receiver are on the $x$-axis, 
we can easily consider the case where they are on the $y$-axis. 

\subsubsection{Mean number of successful receivers}
\label{subsec-exact-mean}
Using Proposition~\ref{prop-cov}, we can also obtain the mean number of successful receivers, 
which is defined as the expected number of vehicles to which the tagged transmitter can transmit. 
The same metric is also considered by Nguyen et al. \cite{Nguy13} under a homogeneous PPP environment. 
%the mean number of  successful receivers 
%%this metric 
%in this paper. 
%also considered by Nguyen et al. \cite{Nguy13} under a homogeneous PPP environment. 
Recall that there are three types of receivers: vehicles in $S_Q$, in $S_{R_x}$, and in $S_{R_y}$. 
Recall also that vehicles transmitting radio waves cannot simultaneously receive information from other vehicles. 
As a result, we obtain the following result. 
\begin{prop}\label{prop-mean}
%If the target transmitter is the $d$-th vehicle from an intersection, 
The mean number of successful receivers $\overline{M}$
for a vehicle distance $d>0$ from an intersection is given by
\begin{align}
\overline{M} 
&=
(1 - \rho) \sum_{i = - n_-}^{n_+}p(|d - il_v|)
+ (1 - \rho_0)
\nonumber
\\
&\times
\left[ 
\lambda_x \!\!\int_{\bbR} p(r) \rmd r
+
\lambda_y \int_{\bbR} p(\sqrt{d^2 + r^2}) \rmd r
\right].
\label{eq-mean-full}
\end{align}
\end{prop}

%\section{Approximate Analysis of Performance Metrics}\label{sec-app}

\section{Approximate Analysis}\label{sec-app}
%\section{Approximate Formulae of Performance Metrics}\label{sec-app}
Although theoretical values of the performance metrics can be obtained 
as in Propositions~\ref{prop-cov} and \ref{prop-mean}, 
they are expressed in non-analytical forms
(especially, due to the terms related to the interference from $S_Q$)
[see (\ref{eq-lap-I_Q})--(\ref{eq-mean-full})]. 
Therefore, it is difficult not only to see the impacts of various parameters on them 
but also to optimize their system parameters because of time-consuming numerical computation. 
% is needed. %integrals and multi-level summations are included. 
%Thus, if we apply any numerical optimization methods to obtain optimal system parameters, a large computational time is required because of these heavy computations in each step of the numerical methods. 
To solve this problem, we attempt to obtain a simple approximation 
for $p(r)$ and $\overline{M}$ that depends only on system parameters
by assuming that the queue length is sufficiently large. 
We then optimize the broadcast rate of vehicles in $S_Q$ (see Section~\ref{subsec-optimize}). 
In accordance with the closed-form approximation, we can solve the optimization problem in a reasonable computational time, and thus, the proposed method can be applied to real-time broadcast rate control for CVS systems. 

\begin{figure}[!t]
\centering
\includegraphics[width=3.5in]{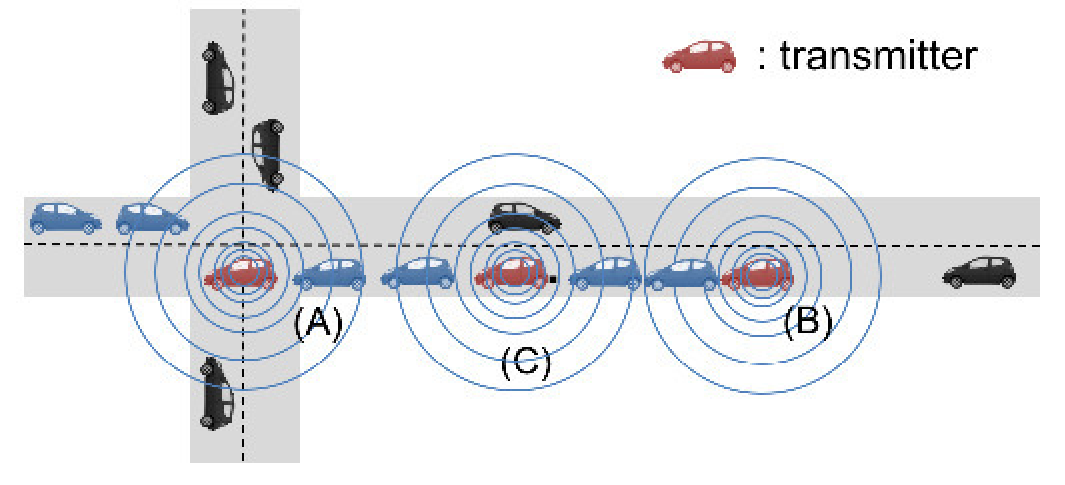}
\caption{Three typical cases considered in Section~\ref{sec-app}:
(A) target transmitter is at intersection, 
(B) at end of queue,
and (C) in  middle of queue. 
}
\label{fig:cases}
\end{figure}

In general, the characteristics of $p(r)$ and $\overline{M}$ depend on the location of 
the tagged transmitter. To obtain approximation formulae, we consider three {\it typical} 
locations of the transmitter instead of considering arbitrary locations: 
the tagged transmitter is in the positive part on the $x$-axis and 
(A) at the intersection, (B) at the end of the queue, and (C) in the middle of the queue (see Figure~\ref{fig:cases}). 
Since a vehicle at (or near) the intersection (case (A)) is affected 
by interferences from both parts ($x$- and $y$-axes) and queues, 
it is expected to have the worst performance. A vehicle near
the end of the queue (case (B)) is said to be in an intermediate state 
of vehicles between the queuing and running segments. 
In case (C), if the queue is sufficiently long, 
the performance can be approximated as vehicles stopping at even intervals on a long 1-d line. 
As shown later, the performance of vehicles at other positions in the queue 
can be estimated by interpolating those in cases (A)--(C) (detailed discussion is in Section~\ref{subsec-other-pos}). 
In addition, we can estimate the other cases where the transmitter is in $S_{R_x}$ or
in $S_{R_y}$ and far from the queue %, however, such cases can be estimated 
by ignoring the effect of the queue and 
considering vehicles homogeneously distributed on a 1-d line.
Therefore, we analyze cases (A)--(C) because they characterize the effect of the intersection. 

%In addition, if a vehicle is in $S_{R_x}$ and far from the end of the queue, 
%its performance can be estimated by considering vehicles homogeneously 
%distributed on a 1-d line. %, i.e., the effect of queues is negligible. 
%%Thus, we do not consider this simple case in this paper. 
%In addition, we can consider %the cases where the transmitter is in $S_{R_x}$ or 
%the case where the transmitter is in $S_{R_y}$ %. However, such cases can be estimated 
%by combining case (A) and a model with a single street. 
%Therefore, we analyze cases (A)--(C) because they characterize the effect of the intersection. 

As we will see later, we can calculate the analytical values of $\calL_{I^X_R}(T r^{\alpha})$ and $\calL_{I^Y_R}(T r^{\alpha})$ in special cases, such as $\alpha \in \bbN$. However, the term $\calL_{I_Q}(T r^{\alpha})$, i.e., the interference from $S_Q$, cannot be expressed in an analytical form even in such cases. Therefore, we mainly focus on giving a closed-form approximation for $\calL_{I_Q}(T r^{\alpha})$ in this paper. The obtained approximate formulae for $\calL_{I_Q}( T r^{\alpha})$ basically hold under conditions in which $\alpha \in \bbN$ and queue lengths $n_+$ and $n_-$ are sufficiently large. 

\subsection{Case (A): Transmitter  at Intersection}\label{subsec-case-A}
We first consider case (A), where the transmitter is at an intersection. 
As mentioned in Section~\ref{subsec-exact-mean}, 
there are three types of receivers: a receiver in  $S_Q$, in  $S_{R_x}$,
and in  $S_{R_y}$.% (see Figure~\ref{fig:cases}). 
%\subsubsection{Probability of successful transmission}\label{subsec-p(r)-case-A}
%
We first provide approximation for the probability of successful transmission
when transmitting to a receiver in  $S_Q$. 
Note that if a receiver is in  $S_Q$
and the $i$-th vehicle from the intersection, 
the communication distance is equal to $i l_v$. 
The main idea is the approximation of $\calL_{I}(s \mid d)$ by considering 
a large queue. By expressing $\log\calL_{I}(s \mid d)$ as 
an infinite series of the interference from each vehicle in the queue
and considering a large queue, we can obtain a closed form approximation
of $p(r)$.  %for any receiver position. 
Detailed explanation for the derivation of the formulae  below
is given in Appendix~\ref{appen-subsec-p(r)-case-A}. 

\medskip
\noindent
{\bf Approximate formulae of $p(r)$ in case (A): }
%\begin{thm}\label{thm-case-A}
Suppose that the transmitter is at an intersection. 
If $(1 - \rho) T \ge 1$\footnote{A typical value of the outage threshold $T$ 
is 10--15 dB (for example, $10\sim15$ dB ($\approx 10\sim35.63$) in IEEE 802.11p). 
In addition, the optimal $\rho$ was often less than $0.4$ in our experiments.  
Thus, this assumption can be considered as valid.
In addition, we can also derive an approximate formula for other cases
using the results in Appendix~\ref{appen-approx}. } and $\alpha \in \bbN$, 
the probability of successful transmission can be approximated as follows. 
(i) If a receiver is the $i$-th vehicle from an intersection, then 
\begin{align}
&p(i l_v)
\approx
%{1 + T \over (1 - \rho)(1 + (1 - \rho)T)}
%\nonumber
%\\
%&~\times
{K(\rho) \over 1 - \rho}
\exp
\left[\left(
2 \xi_{\alpha, T}(\rho)
- \rho_0 \left(\lambda_x C^X_{\alpha, T} + \lambda_y C^Y_{\alpha, T}\right)l_v
\right)i
\right], 
%\nonumber
\label{eq-prob-app-case-A}
\end{align}
where
\begin{align}
\xi_{\alpha, T}(\rho)
&=
(\alpha + \kappa_{1,\alpha} - \kappa_{2,\alpha})
 ((1 - \rho)^{1/\alpha}- 1)T^{1/\alpha},
% \nonumber
\label{eq-def-xi}
\\
\kappa_{1, \alpha} &= \alpha \sum_{k=1}^{\infty}{(-1)^{k+1} \over \alpha k - 1},~~
\kappa_{2, \alpha} =  \alpha \sum_{k=1}^{\infty}{(-1)^{k+1} \over \alpha k + 1},
\label{eq-def-kappa}
%\nonumber
\end{align}
and 
{\small
\begin{align}
&K(\rho) = {1 + T \over 1 + (1 - \rho)T}, 
\label{eq-def-K(rho)}
\\
%&C_Q = {1 + T \over (1 - \rho)(1 + (1 - \rho)T)},
%\\
&C^X_{\alpha, T} = 
2 T^{1 \over \alpha}
{\pi \over \alpha} \cosec\left({\pi  \over \alpha}\right),
%\int_{-\infty}^{\infty}{\rmd x \over x^{\alpha} + 1},
%\left\{
%\begin{array}{ll}
%{T^{1 \over 4} \over \sqrt{2}}
%, & \alpha = 4,\\
%\sqrt{T}, & \alpha = 2,
%\end{array}
%\right.
\quad 
C^Y_{\alpha, T} = \int_{\bbR} {T \rmd y \over (y^2 + 1)^{\alpha \over 2} + T},
%\left\{
%\begin{array}{ll}
%{\sqrt{T} \sqrt{\sqrt{1 + T} - 1}
%\over 
%\sqrt{2} \sqrt{1 + T}}, & \alpha = 4,\\
%{ T \over \sqrt{ 1 + T}}, & \alpha = 2,
%\end{array}
%\right.
\label{eq-def-C_T}
%\nonumber
\end{align}
}
(ii) if a receiver is in  $S_{R_x}$ at distance $r>0$ from the intersection, then
{\small
\begin{equation}
p(r)
\approx
K(\rho)
%\nonumber
%\\
%&\times&
\exp
\left[\left(
{2 \xi_{\alpha, T}(\rho) \over l_v}
- \rho_0 \left(\lambda_x C^X_{\alpha, T} + \lambda_y C^Y_{\alpha, T}\right)
\right)r
\right], \qquad 
%\nonumber
\label{eq-prob-app-case-A-r-x}
\end{equation}
}
%
%with 
%%
%\[
%K(\rho) = {1 + T \over 1 + (1 - \rho)T}, 
%\]
%%
and (iii) if a receiver is in  $S_{R_y}$ at distance $r>0$ from the intersection, then
{\small
\begin{align}
p(r)
&\approx
{K(\rho) \over (1 - \rho)^{2 r }}
\exp
\left[\left(
%{2 \xi_{\alpha, T}(\rho) \over l_v} 
-{2\rho \over (\alpha + 1)(1- \rho)Tl_v}
\right.\right.
\nonumber
\\
&~~
%-
%\exp
%\left[\left(
%{2 \xi_{\alpha, T}(\rho) \over l_v} 
\left.\left.
+{2 \xi_{\alpha, T}(\rho) \over l_v} 
%{2\rho \over (\alpha + 1)(1- \rho)Tl_v}
- \rho_0 \left(\lambda_x C^X_{\alpha, T} + \lambda_y C^Y_{\alpha, T}\right)
\right)r
\right]. 
\label{eq-prob-app-case-A-r}
%\nonumber
\end{align}
}

%
%\end{thm}
%
%
\begin{rem}
If $\alpha = 2, 4$, $C_{\alpha, T}^Y$ can be computed as follows. 
\begin{eqnarray*}
%&{}&C^X_{\alpha, T} = 
%{\pi \over \alpha} \csc\left({\pi  \over \alpha}\right),\quad \alpha \in \bbN,
%\nonumber
%\\
%&{}&
%C^Y_{\alpha,T} = 
%\left\{
%\begin{array}{ll}
%{ T \over \sqrt{ 1 + T}}, & \alpha = 2,\\
%{\sqrt{T} \sqrt{\sqrt{1 + T} - 1}
%\over 
%\sqrt{2} \sqrt{1 + T}}, & \alpha = 4.
%\end{array}
%\right.
C^Y_{2, T} = { T \over \sqrt{ 1 + T}},\qquad
C^Y_{4, T} = 
{\sqrt{T} \sqrt{\sqrt{1 + T} - 1}
\over 
\sqrt{2} \sqrt{1 + T}}. 
\end{eqnarray*}
%
%\[
\end{rem}

\begin{rem}
The above approximate formulae %corresponding to cases (i)--(ii)
(\ref{eq-prob-app-case-A}), (\ref{eq-prob-app-case-A-r-x}),
 and (\ref{eq-prob-app-case-A-r}) 
suggest that, in our approximation,
the probability of successful transmission decreases geometrically with 
the distance to receivers, and the decay rate is 
determined by only system parameters. 
In addition, if the parameters $\alpha$ and $T$ that depend on a system or environment
are given in advance, $\kappa_{1,\alpha}$, $\kappa_{2,\alpha}$, 
$C^X_{\alpha, T}$ and $C^Y_{\alpha, T}$ can be regarded as constant. 
\end{rem}

The approximate formulae~(\ref{eq-prob-app-case-A}), (\ref{eq-prob-app-case-A-r-x}),
 and (\ref{eq-prob-app-case-A-r}) suggest that, in our approximation,
the probability of successful transmission decreases geometrically with 
the distance to receivers. For example, if a receiver is in  $S_Q$, 
then the geometric decay rate is equal to 
\[
\exp\left(
2 \xi_{\alpha, T}(\rho)
- \rho_0 \left(\lambda_x C^X_{\alpha, T} + \lambda_y C^Y_{\alpha, T}\right)l_v
\right),
\]
which is determined by only system parameters and can be easily computed
using (\ref{eq-def-xi})--(\ref{eq-def-C_T}). The same applies to 
the case where a receiver is in  $S_{R_x}$ or  $S_{R_y}$.

%\subsubsection{Mean number of successful receivers}\label{subsec-mean-A}

From the results in the previous section, 
 we can approximate the mean number of successful receivers. 
Since the approximate formulae of $p(r)$ are expressed in a geometric form, 
we can also obtain a closed-form approximation for $\overline{M}$. 
Let $\overline{M}_Q(\rho)$, $\overline{M}_{R_X}(\rho)$, and $\overline{M}_{R_Y}(\rho)$
denote the mean numbers of successful receivers in  $S_Q$, $S_{R_x}$, and $S_{R_y}$,
respectively. 
First, applying (\ref{eq-prob-app-case-A})
%statement (i) of Theorem~\ref{thm-case-A} 
to Proposition~\ref{prop-mean}  
and considering sufficiently large
$n_+$ and $n_-$, we obtain
\begin{eqnarray}
\overline{M}_Q(\rho) \approx
2 (1 - \rho) \sum_{i=1}^{\infty} p(i l_v).
\nonumber
%\approx
%{1 + T \over 1 + (1 - \rho)T}
%} &&
%\nonumber
%\\
%%&\approx&
%&{}&
%\times
%{2 \rho \exp\left( 2\xi_{\alpha, T}(\rho) - \pi\rho_0(\lambda_x C^X_{\alpha, T} + \lambda_y C^Y_{\alpha, T}) l_v\right)
%\over
% 1 - \exp\left( 2\xi_{\alpha, T}(\rho) - \pi\rho_0(\lambda_x C^X_{\alpha, T} + \lambda_y C^Y_{\alpha, T}) l_v  \right)},
% \quad
%\nonumber
%\\
%\label{eq-M_Q-case-A}
\end{eqnarray}
Similar to the above, from (\ref{eq-prob-app-case-A-r-x}) and (\ref{eq-prob-app-case-A-r}),
%Similar to the above, from the statements (ii) and (iii) of Theorem~\ref{thm-case-A}, 
we can approximate $\overline{M}_{R_X}(\rho)$ and $\overline{M}_{R_Y}(\rho)$ as follows.
Thus, under the same conditions as in $p(r)$, we obtain their approximation as follows.

\noindent
{\bf Approximate formulae of $\overline{M}$ in case (A): }
{\small
\begin{align}
&\overline{M}_Q(\rho) 
\approx
{2 K(\rho) \exp\left( 2\xi_{\alpha, T}(\rho) - \rho_0(\lambda_x C^X_{\alpha, T} + \lambda_y C^Y_{\alpha, T}) l_v\right)
\over
 1 - \exp\left( 2\xi_{\alpha, T}(\rho) -\rho_0(\lambda_x C^X_{\alpha, T} + \lambda_y C^Y_{\alpha, T}) l_v  \right)},
%\nonumber
\label{eq-M_Q-case-A}
\\
&\overline{M}_{R_X}(\rho) 
%=
%2 (1 - \rho_0) \lambda_x \int_{0}^{\infty} p(r) \rmd r
%\nonumber
%\\
%&\qquad
\approx
{2 K(\rho) (1 - \rho_0) \lambda_x l_v
\over
2\xi_{\alpha, T}(\rho)- \rho_0(\lambda_x C^X_{\alpha, T} + \lambda_y C^Y_{\alpha, T}) l_v },
\label{eq-M_R-case-A-x}
%\nonumber
\\
%\end{eqnarray}
%%
%\begin{eqnarray}
&\overline{M}_{R_Y}(\rho) 
%&=&
%2 (1 - \rho_0) \lambda_y \int_{r=0}^{\infty} p(r) \rmd r
%\nonumber
%\\
\approx
2 (1 - \rho_0) \lambda_y l_v K(\rho)
%\left({ 1 + T \over 1 + (1 - \rho)T  } \right)
\left[
2\xi_{\alpha, T}(\rho)- 2 \log(1 - \rho) 
\right.
\nonumber
\\
&\qquad
-
\left.
%{2 K(\rho) \over \alpha + 1}
{2\rho \over (\alpha + 1)(1- \rho)T}
- \rho_0(\lambda_x C^X_{\alpha, T} + \lambda_y C^Y_{\alpha, T}) l_v
\right]^{-1}.
%\nonumber
\label{eq-M_R-case-A}
\end{align}
}

\subsection{Case (B): Transmitter at  End of Queue}\label{subsec-case-B}
We next consider case (B), where the transmitter is at the end 
of the queue. In this case, the transmitter is far from the $y$-axis 
due to the queueing segment. Therefore, the interferences from the 
vehicles in  $S_{R_y}$ and the receivers in  $S_{R_y}$ are 
both negligible. 
This case can be divided into three sub-cases: 
a receiver is in (i) $S_Q$,  in (ii) $S_{R_x}$ in the negative direction, 
or (iii) $S_{R_x}$ in the positive direction,  
i.e., the left-hand side of the transmitter 
or the right-hand side (see Figure~\ref{fig:cases}). 
%\subsubsection{Probability of successful transmission}\label{subsec-p(r)-case-B}
%
Since the interference from the vehicles in  $S_{R_y}$ 
is relatively much smaller than that from  $S_Q$ and $S_{R_x}$, 
%regardless of the position of a receiver. 
the term $\calL_{I_Y}(T r^{\alpha})$ is negligible, i.e.,
\begin{equation}
p(r) \approx \calL_{I_Q}( T r^{\alpha})\calL_{I_X^R}( T r^{\alpha}).
\label{eq-app-p(r)-Lap_X-Q}
\end{equation}
We then have the following results, in which $p(r)$ also decreases geometrically as 
$r$ increases; however, the decay rate is different from that  in case (A).
Detailed explanation for the derivation of the formulae below
is given in Appendix~\ref{appen-subsec-p(r)-case-B}. 
%Theorem~\ref{thm-case-A}. 
%

\medskip
\noindent
{\bf Approximate formulae of $p(r)$ in case (B): }
%\begin{thm}\label{thm-app-case-B-i}
Suppose that the transmitter is at the end of the queue and $(1 - \rho) T \ge 1$. 
If $\alpha \in \bbN$, the probability of successful transmission 
can be approximated as follows.
(i) If a receiver is in $S_Q$ and the $i$-th vehicle from the end of the queue, then
{\small
\begin{align}
p(i l_v) 
&\approx
K(\rho)
%{ e^{\rho \over 2 (1 - \rho)T}\over \sqrt{1 - \rho}}
e^{\rho \over 2 (1 - \rho)T}
(1- \rho)^{i - {1 \over 2}} 
%{1 + T \over 1 + (1 - \rho)T}
e^{
\left(
\beta_{\alpha, T}(\rho) - \rho_0 \lambda_x C^X_{\alpha, T} l_v 
\right)
i}
%\nonumber
%\\
%&{}
%~~~\times
%\exp
%\left[
%\left(
%\beta_{\alpha, T}(\rho) - \rho_0 \lambda_x C^X_{\alpha, T} l_v 
%\right)
%i\right],\qquad
\label{eq-app-case-B-i}
\end{align}
}
where
\begin{equation}
\beta_{\alpha, T}(\rho) = \xi_{\alpha, T}(\rho) + {\rho \over (1- \rho)(\alpha + 1)T} ,
\label{eq-def-beta}
%\nonumber
\end{equation}
(ii) if a receiver is in $S_{R_x}$ in the negative part
and at distance $r > 0$ from the end of the queue, then
{\small
\begin{align}
p(r) 
&\approx
K(\rho)
 e^{\rho \over 2 (1 - \rho)T}(1- \rho)^{{r \over l_v} + {1 \over2 }} 
 e^{\left(
{\beta_{\alpha, T}(\rho)  \over l_v} - \rho_0 \lambda_x C^X_{\alpha, T} 
\right)},
% {1 + T \over 1 + (1 - \rho)T}
%\nonumber
%\\
%&
%~~~\times 
%\exp
%\left[
%\left(
%{\beta_{\alpha, T}(\rho)  \over l_v} - \rho_0 \lambda_x C^X_{\alpha, T} 
%\right)
%r\right],\qquad
\label{eq-app-case-B-i-x}
\end{align}
}
and (iii) if a receiver is in  $S_{R_x}$ in the positive part
and at distance $r > 0 $ from the end of the queue, then
{\small
\begin{align}
p(r) 
&\approx
\sqrt{K(\rho)}
%\sqrt{{ 1 + T \over 1 + (1 - \rho)T }}
(1 - \rho)^{-{r \over l_v}}
\exp
\left[\left(
{\xi_{\alpha, T}(\rho) \over l_v}
%{ \over  \sqrt{1 - \rho}}
\right.
\right.
\nonumber
\\
&{}~~~
\left.
\left.
%\exp
%\left[\left(
%{\xi_{\alpha, T}(\rho) \over l_v}
-
{\rho \over (\alpha + 1)(1- \rho)Tl_v}
-  \rho_0 \lambda_x C^X_{\alpha, T}
\right) r\right].
\label{eq-app-case-B-r}
%\nonumber
%\\
\end{align}
}

%\end{thm}
%

%\subsubsection{Mean number  of successful receivers}
%
Similar to case (A) considered in Section~\ref{subsec-case-A}, 
the approximate formulae presented in the previous section %Section~\ref{subsec-p(r)-case-B}
%Theorem~\ref{thm-app-case-B-i}
are in geometric forms. This fact again enables us
to obtain the closed-form approximation $\overline{M}(\rho)$. 
Recall here that $\overline{M}_{R_Y}(\rho)$ is negligible in this case 
due to the distance between the transmitter and the $y$-axis. 
Thus, by using (\ref{eq-app-case-B-i}), (\ref{eq-app-case-B-i-x}), (\ref{eq-app-case-B-r}),
%Theorem~\ref{thm-app-case-B-i} 
and Proposition~\ref{prop-mean}, 
we can approximate $\overline{M}_Q(\rho)$ and $\overline{M}_{R_X}(\rho)$ as below. 

\noindent
{\bf Approximate formulae of $\overline{M}$ in case (B): }
{\small
\begin{align}
&\overline{M}_Q(\rho) 
\approx
\sqrt{1 - \rho} \exp\left( {\rho \over 2 (1- \rho)T }\right) 
K(\rho)
%{1 + T \over 1 + (1 - \rho)T}
%} \qquad&&
\nonumber
\\
&~~~\times
{
 \exp\left( \beta_{\alpha, T}(\rho) - \rho_0 \lambda_x C^X_{\alpha, T} l_v \right)
\over
1- (1 - \rho) \exp\left( \beta_{\alpha, T}(\rho) -  \rho_0 \lambda_x C^X_{\alpha, T}  l_v \right)
},\qquad
\label{eq-M_Q-case-B}
\\
%\end{eqnarray}
%%
%\begin{eqnarray}
&\overline{M}_{R_X}(\rho)
\approx
\sqrt{1 - \rho} \exp\left( {\rho \over 2 (1- \rho)T}\right) 
K(\rho) % {1 + T \over 1 + (1 - \rho)T}
% } &&
 \nonumber
 \\
&\times
{
 (1 - \rho_0)\lambda_x l_v 
\over
\beta_{\alpha, T}(\rho) + \log(1 - \rho) - \rho_0 \lambda_x C^X_{\alpha, T}  l_v}
+
 (1 - \rho_0) \lambda_x l_v\sqrt{K(\rho)}
%\sqrt{{ 1 + T \over 1 + (1 - \rho)T  }}
\nonumber
\\
%&{}&
%\nonumber
%\\
&
\times
\left[
\xi_{\alpha, T}(\rho)
-
\log(1 - \rho) 
-
{\rho \over (\alpha + 1)(1- \rho)T}
- \rho_0\lambda_x C^X_{\alpha, T} l_v
\right]^{-1}.
%+ 
%{1 \over \sqrt{1 - \rho} }
%{(1 - \rho_0)\lambda_x l_v 
%\over
%\beta_{\alpha, T}(\rho) - \log(1 - \rho) - \pi \rho_0 \lambda_x C^X_{\alpha, T}  l_v}.
%\nonumber
%\\
%=nonumber
\label{eq-M_R-case-B}
\end{align}
}
\subsection{Case (C): Transmitter in  Middle of Queue}\label{subsec-case-C}
Finally, we consider case (C), where the transmitter is in the middle of the queue. 
%\subsubsection{Probability of successful transmission}\label{subsec-p(r)-case-C}
%\subsubsection{Probability of successful transmission}\label{subsec-p(r)-case-C}
As well as case (B), if the queue is sufficiently long, then we can neglect the interference from 
 $S_{R_y}$ and $\overline{M}_{R_Y}(\rho)$. 
Thus, we approximate this case by considering vehicles queuing at even intervals on a single street
with infinite length, i.e., a single infinite queue. 
Under this assumption, we can obtain the approximate formulae for this case 
by simply removing the effect of the interference from vehicles on the $y$-axis
in the results in Section~\ref{subsec-case-A}. 
%of Theorem~\ref{thm-case-A}. 
Thus, substituting $\lambda_y = 0$ into (\ref{eq-prob-app-case-A}) and 
(\ref{eq-prob-app-case-A-r-x}), we can
immediately obtain the following. 

\medskip
\noindent
{\bf Approximate formulae of $p(r)$ in case (C): }
%\begin{thm}\label{thm-case-C}
Suppose that the transmitter is in the middle of the queue. 
If $(1 - \rho) T \ge 1$ and $\alpha \in \bbN$, 
the probability of successful transmission can be approximated as follows.
(i) If a receiver is the $i$-th vehicle from the transmitter, then 
\begin{eqnarray}
p(i l_v)
&\approx&
%{1 + T \over (1 - \rho)(1 + (1 - \rho)T)}
{K(\rho) \over 1 - \rho}
%\nonumber
%\\
%&{}&
%\times
\exp
\left[\left(
2 \xi_{\alpha, T}(\rho)
- \rho_0 \lambda_x C^X_{\alpha, T} l_v
\right)i
\right],
\label{eq-prob-app-case-C}
\end{eqnarray}
and (ii) if a receiver is in $S_{R_x}$ at distance $r$ from the intersection, then
\begin{equation}
p(r)
\approx
%{1 + T \over 1 + (1 - \rho)T} 
K(\rho)
%\nonumber
%\\
%&\times&
\exp
\left[\left(
{2 \xi_{\alpha, T}(\rho) \over l_v}
- \rho_0 \lambda_x C^X_{\alpha, T}
\right)r
\right].
\label{eq-prob-app-case-C-x}
\end{equation}
%
%
%\end{thm}
%

%\medskip

%\subsubsection{Mean number of successful receivers}\label{subsec-mean-C}

As mentioned in the above, the number of the successful receivers in  $S_{R_y}$ is 
relatively small in this case. 
Therefore, it is sufficient to consider receivers in  $S_Q$ and $S_{R_x}$. 
In a similar way to the
derivation of (\ref{eq-M_Q-case-A}), 
we also easily obtain an approximation for $\overline{M}_Q(\rho)$ and $\overline{M}_{R_X}(\rho)$
by substituting $\lambda_y = 0$ into  (\ref{eq-M_Q-case-A}) and (\ref{eq-M_R-case-A-x}), respectively. 

\noindent
{\bf Approximate formulae of $\overline{M}$ in case (C): }
\begin{eqnarray}
%\lefteqn{
%\overline{M}_Q(\rho)
%} &&
%\nonumber
%\\
\overline{M}_Q(\rho)
&\approx&
%{1 + T \over 1 + (1 - \rho)T}
{2 \rho K(\rho) \exp\left( 2\xi_{\alpha, T}(\rho) - \rho_0 \lambda_x C^X_{\alpha, T} l_v\right)
\over
 1 - \exp\left( 2\xi_{\alpha, T}(\rho) - \rho_0 \lambda_x C^X_{\alpha, T} l_v  \right)},
\nonumber
\\
\label{eq-M_Q-case-C}
\end{eqnarray}
Similar to the above, from (\ref{eq-prob-app-case-A-r-x}) and (\ref{eq-prob-app-case-A-r}),
%Similar to the above, from the statements (ii) and (iii) of Theorem~\ref{thm-case-A}, 
we can approximate $\overline{M}_{R_X}(\rho)$ and $\overline{M}_{R_Y}(\rho)$ as follows.
\begin{eqnarray}
%\lefteqn{
\overline{M}_{R_X}(\rho) 
%} &&
%\nonumber
%\\
&\approx&
%{1 + T \over 1 + (1 - \rho)T}
{2 K(\rho) (1 - \rho_0) \lambda_x l_v
\over
2\xi_{\alpha, T}(\rho)- \rho_0 \lambda_x C^X_{\alpha, T} l_v }.
\quad
%\nonumber
\label{eq-M_R-case-C-x}
\end{eqnarray}
\section{Broadcast Rate Optimization}\label{subsec-optimize}
We next consider the optimization of the broadcast rate of vehicles in the $S_Q$ 
on the basis of the approximate formulae
%of the performance metrics 
presented in Section~\ref{sec-app}. We assume that vehicles can determine their status 
(i.e., queuing or running) by tracking their speed. 
%In addition, the detailed position of a vehicle in the queue 
%%(i.e., (A), (B), or (C)) 
%can be determined by calculating the distance from an intersection from location information such as GPS. 
If the vehicles in $S_Q$ transmit with a high broadcast rate, 
then they have higher interference than those in $S_R$ due to the congestion of vehicles at the intersection. 
%Therefore, $p(r)$ of the transmitter in $S_Q$ becomes lower than that of the transmitter in $S_R$. 
However, 
%there are more vehicles near the intersection than in $S_R$, and thus, the transmitter in $S_Q$ has more vehicles to which to transmit. In addition, 
if a vehicle transmits with a high broadcast rate, it has more chance to successfully transmit to its neighbors
(to be discovered by the neighbors). 
Thus, by carefully choosing the broadcast rate of the vehicles in $S_Q$, we can mitigate the interference 
and improve the performance of the V2V communication. To characterize and balance this relationship, 
we consider {\it the mean number of successful transmissions per unit time}, which is equal to 
\begin{eqnarray}
D(\rho) &=& \rho \overline{M}(\rho).
%\label{eq-def-D-rho}
\nonumber
\end{eqnarray}
%
%Note here that we directly optimize the value $\rho$, not $\theta$, because $\rho = \theta L$ and $L$ is assumed to be constant. 
In the CVS systems, vehicles periodically transmit a packet so that other vehicles know
their positions, i.e., they can be discovered by other vehicles. 
Therefore, this metric can be considered as the number of {\it discoveries} for a typical transmitter 
per unit time and a key performance metric in V2V broadcast communications. 
%As we mentioned a similar metric is also considered by Nguyen et al. \cite{Nguy13}
We can consider other metrics, such as probability of successful transmission to the nearest vehicle~\cite{Kimu18}, 
however, to focus on the performance of the broadcast communication, we consider this metric. 
Using $D(\rho)$, we consider the optimization problem
\begin{equation}
\rho_{\ast} = \underset{0 \le \rho \le 1}{\arg\!\max}~D(\rho). 
%\label{eq-opt}
\nonumber
\end{equation}
By numerically solving the above problem, we can obtain the optimal broadcast rate that maximizes $D(\rho)$. 
%the mean number of successful transmissions per unit time. 
%In addition, by substituting the approximate formulae of $\overline{M}_Q(\rho)$, $\overline{M}_{R_X}(\rho)$, and $\overline{M}_{R_Y}(\rho)$ [shown in (\ref{eq-M_Q-case-A})--(\ref{eq-M_R-case-A}), (\ref{eq-M_Q-case-B}), and (\ref{eq-M_R-case-B})] with (\ref{eq-def-D-rho}), we can consider the optimization problem of the broadcast rate in cases (A)--(C). 
Recall that the values from exact analysis shown in Propositions~\ref{prop-cov} and \ref{prop-mean} 
require time-consuming numerical computation, and thus
%include numerical integrals and multi-level summations, 
%the calculation of the exact value of $D(\rho)$ also requires a large computational cost. 
the optimization of $D(\rho)$ becomes much more time-consuming because of iterative computation in numerical optimization methods. 
%such a heavy computation is needed in each step of numerical optimization methods. 
However, by using the closed-form approximation of $D(\rho)$, we can compute the optimal $\rho_{\ast}$ in a reasonable computational time. 
%Fortunately, since $D(\rho)$ becomes convex in a realistic situation such as $\rho < 0.5$ (see Section~\ref{}), we can easily find
%Because we found that $D(\rho)$ and $M(\rho)$ are mostly insensitive to $n_+$ and $n_-$, we adopted a long queue assumption in our approximation. 
%
Indeed, if we assume that $\alpha$, $T$, and $\rho_0$ are given in advance, $\rho_{\ast}$ can be determined by only $\lambda_x$ and $\lambda_y$. This fact suggests that if we prepare a look-up table in our vehicles that describes the optimal broadcast rate corresponding to each value of $\lambda_x$ and $\lambda_y$, 
%we can immediately optimize the broadcast rate by just obtaining the current traffic intensity in the running segments. This fact leads to the possibility of 
we can control the broad cast rate in (near-)optimal real-time manner. 

Although $D(\rho)$ and the optimal $\rho$ depends on the positions of the transmitters (i.e., cases (A)--(C)), 
%we adopted a long queue assumption in our approximation. 
we found that if the intensity in $S_R$ is not very high, $\rho_{\ast}$ 
is almost insensitive to the cases (A)--(C) in our numerical examples. 
Thus, we can obtain near-optimal broadcast rate regardless of 
the position of the tagged transmitter. % by considering cases (A)--(C). 
We also found that they are mostly insensitive to $n_+$ and $n_-$, 
which indicates that our 
large queue assumption is valid for the broadcast rate optimization
(see Section~\ref{subsec-eval-opt}).

%%%%%%%%%%%%%%%%%%%%%%
It should be noted that although the vehicles in queueing segments do not
move, it is important to continue to send packets so that vehicles can track
their status (i.e., positions). In addition, data size flying in vehicular 
networks may become much larger in the future (for example, in-vehicle video). 
Therefore, we consider the above scenario in which the mean number of
successfully transmitted packet is optimized. 
%%%%%%%%%%%%%%%%%%%%%%%

\section{Numerical Examples}\label{sec-experiments}

In this section, we provide several numerical examples. 
We first show the results for the performance metrics
$p(r)$ and $\overline{M}(\rho)$ and 
evaluate our approximation %in cases (A)--(C)
in Sections~\ref{sec-app}. 
%\ref{subsec-case-A}--\ref{subsec-case-C}
%with both simulation and our exact model results. 
We then discuss our broadcast rate optimization method. 
%presented 
%in Section~\ref{subsec-optimize}. 
Finally, we investigate the performance of vehicles at other locations
by interpolating or extrapolating the results for cases (A)--(C).
%Finally, 

Before we move on to the numerical results, we will explain the parameters
used in the examples. 
The interval of vehicles $l_v$ was fixed to 
$6$~[m] and $\alpha = 4$ in all examples. By considering realistic
settings, we chose $\rho_0 = 0.1$ and $T = 15$~[dB]. 
In addition, $\lambda \triangleq \lambda_x = \lambda_y = 35$ [km$^{-1}$]
and $n_+ = n_- = N$. 
In each round of the numerical simulation, 
we first set vehicles in $S_Q$s and those in $S_R$s on the basis 
of PPPs on roads 10~km long. 
The vehicles are assumed to be stationary
during the simulation and are static. 
We then calculated the SIR of each receiver by randomly sampling 
the value of fading. We conducted 10,000 numerical simulations 
for each graph. 
Moreover, all error-bars in the graphs in this paper represent 95\% confidence intervals.

\begin{figure}[!t]
\centering
\includegraphics[width=1.7in]{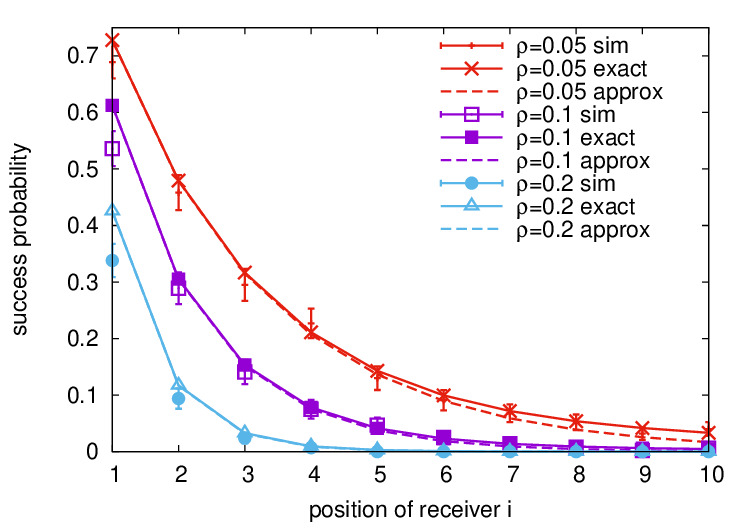}~
\includegraphics[width=1.7in]{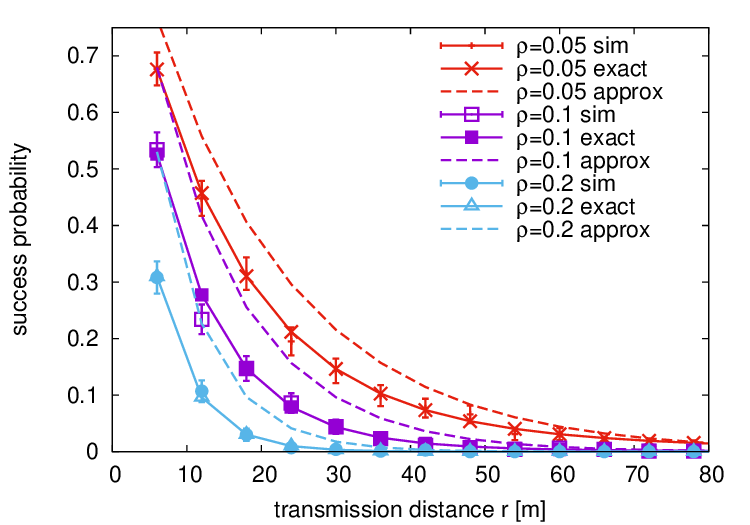}
\caption{Comparison of values of $p(r)$ in case (A) from simulation/exact/approximate analysis with different $\rho$ when $N = 25$ and receiver is $i$-th vehicle from intersection in $S_Q$ (left) and $S_{R_y}$ (right). Each point with error bar (sim), solid line (exact), and dashed line (approx) represent simulation results, our exact model results, and our approximation results. }
\label{fig:exp-1-case-A}
\vskip -7pt
\end{figure}

\begin{figure}[!t]
%\centering
\begin{minipage}[t]{0.48\hsize}
%\centering
%\includegraphics[width=1.7in]{imgs/exp_1_case_B.eps}
\includegraphics[width=1.7in]{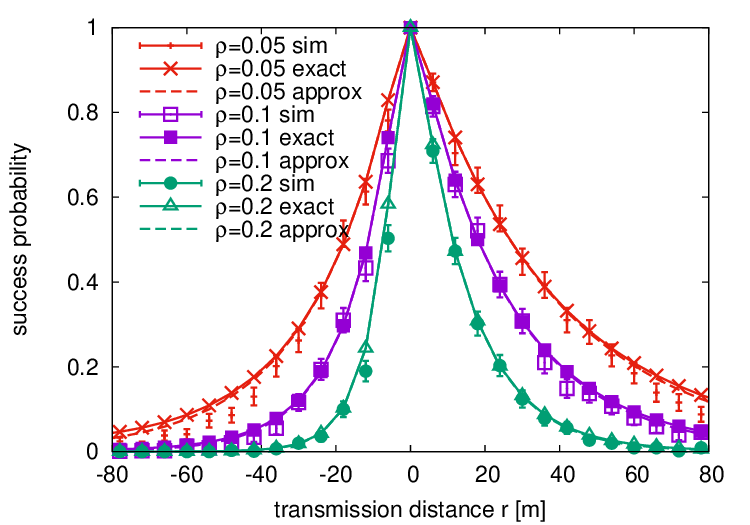}
\caption{Comparison of values of $p(r)$ in case (B) from simulation/exact/approximate analysis
with different $\rho$ when $N = 25$.}
\label{fig:exp-1-case-B}
\end{minipage}
~
\begin{minipage}[t]{0.48\hsize}
%\centering
%\includegraphics[width=1.7in]{imgs/exp_1_case_C.eps}
\includegraphics[width=1.7in]{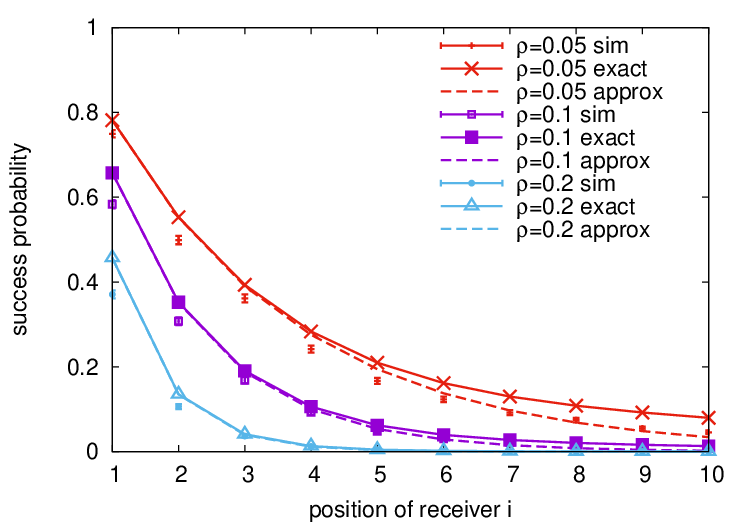}
\caption{Comparison of values of $p(r)$ in case (C) from simulation/exact/approximate analysis
with different $\rho$ when $N = 25$.}
\label{fig:exp-1-case-C}
\end{minipage}
\vskip -7pt
\end{figure}

\begin{figure*}[!t]
\centering
\includegraphics[width=2.0in]{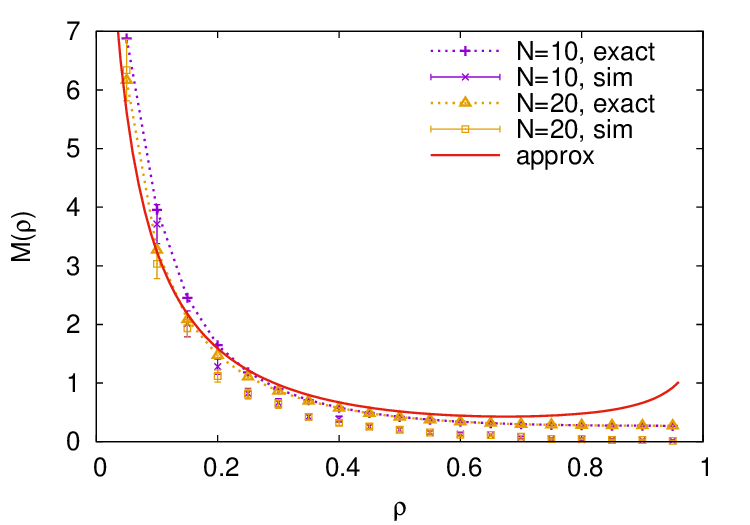}~~~
\includegraphics[width=2.0in]{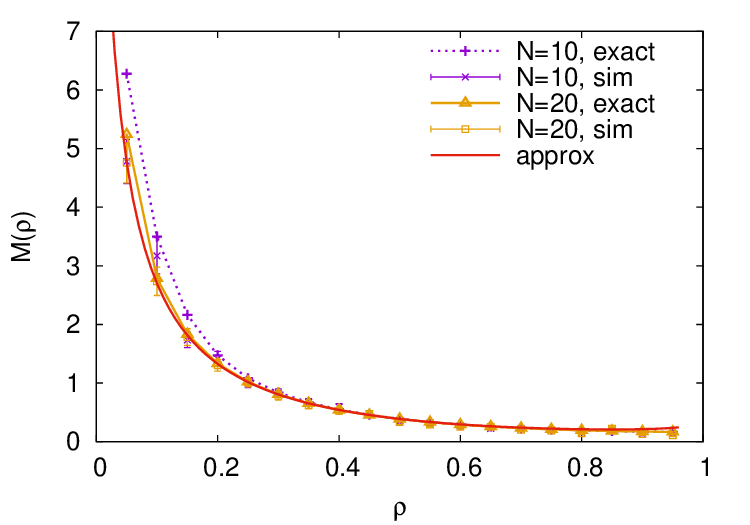}~~~
\includegraphics[width=2.0in]{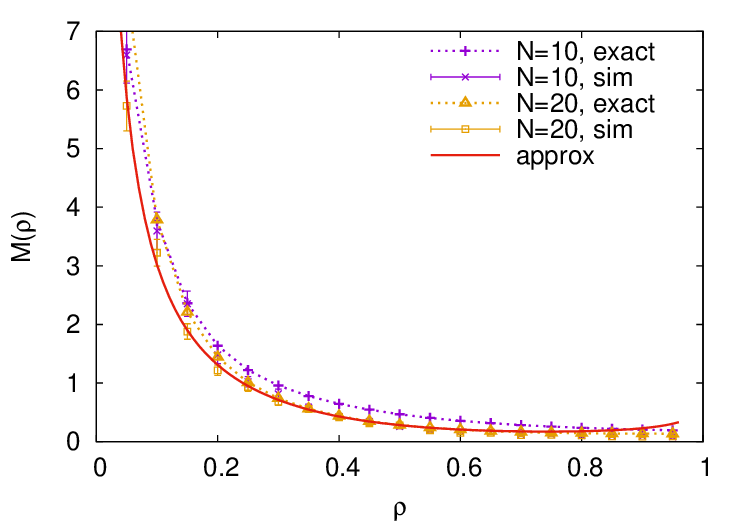}
\caption{Comparison of values of $\overline{M}(\rho)$ from simulation/exact/approximate analysis with different $\rho$ and $N$. 
Left, middle, and right figures correspond to cases (A), (B), and (C). }
\label{fig:exp-2-case}
\vskip -7pt
\end{figure*}

\subsection{Evaluation of Performance Metrics}

We first provide the numerical results for the performance metrics $p(r)$ and $\overline{M}(\rho)$ and evaluate the accuracy of our approximate formulae for them. 
Figure~\ref{fig:exp-1-case-A} compares the simulation results and 
the exact and approximate values of $p(r)$ in case (A), i.e., the case where the transmitter is at the intersection
(see Section~\ref{subsec-case-A}). 
The left graph corresponds to the case where the receiver is in the $S_Q$ and the $i$-th vehicle from the intersection. In addition, the right graph corresponds to the case where the receiver is in $S_{R_y}$, and the horizontal axis represents the transmission distance. 
We calculated the values from exact analysis using (\ref{eq-cov-full}) and those from approximate analysis using (\ref{eq-prob-app-case-A}) [left graph] and (\ref{eq-prob-app-case-A-r}) [right graph]. We can see from the left graph that 
%$p(r)$ geometrically decreased, and 
if $\rho$ increases, $p(r)$ also decreases due to higher interference from vehicles in $S_Q$. We can also see that our approximate formulae fitted well to the results from simulation and exact analysis in all cases and the error became larger when $i$ was larger. 
Since we assume that $N$ is sufficiently large in our approximation, if the distance from the receiver to the end of the queue is closer, then the approximation error becomes large. From the right graph, we can find that the approximate formulae took higher values than the theoretical results and the error increased subject to $\rho$. 
The reason for this is that we approximate the Euclidean distance from the receiver in $S_{R_y}$ to the transmitter at the intersection by the Manhattan distance (see (\ref{eq-L_Q_ii-def-2}) in Appendix). Since the Manhattan distance is larger than the Euclidean distance, the interference became smaller and $p(r)$ became larger than in the simulation and exact analysis. 
In addition, the larger $\rho$ suggests that there were greater impacts from the interference from the vehicles in $S_Q$. Therefore, the errors increased subject to $\rho$. Although the right graph contains larger errors than the left one, we could obtain a rough estimation for $p(r)$. Indeed, we later determined that the errors could be negligible when considering $\overline{M}(\rho)$ (see Fig.~\ref{fig:exp-2-case}). 
Similarly, Figs.~\ref{fig:exp-1-case-B} and \ref{fig:exp-1-case-C} show the same results in cases (B) and (C) where the transmitter is at the end of the queue and case (C) in the middle of the queue. 
%of our approximation and the simulation and exact analysis for $p(r)$ in case (B) where the transmitter is at the end of the queue. 
The horizontal axis in Fig.~\ref{fig:exp-1-case-B} represents the distance to the receiver where the positive (resp. negative) part corresponds to the vehicles in the right-hand (resp. the left-hand) side
, i.e., in $S_{R_x}$ (resp. the left-hand side, i.e., in $S_Q$) of the transmitter. 
We used (\ref{eq-app-case-B-i}) and (\ref{eq-app-case-B-r}) for the approximate values. 
The figures show that our approximate formulae achieved quite small errors in all cases. 
In addition, we can see from  Fig.~\ref{fig:exp-1-case-B} that if $\rho$ is smaller, the results on the positive and negative parts become closer because the interference from the $S_Q$ decreases. %This is because the interference from the $S_Q$ decreases as $\rho$ decreases. 
%Figure~\ref{fig:exp-1-case-C} shows the numerical results for case (C), where the transmitter is in the middle of the queue. 
%The values of exact analysis in (\ref{eq-cov-full}) were approximated by using (\ref{eq-prob-app-case-C}). The figure shows a similar tendency to case (A) (see Figure~\ref{fig:exp-1-case-A}) because the results can be computed very similarly. 

We next show the results for $\overline{M}(\rho)$. Figure~\ref{fig:exp-2-case} compares the simulation results and the exact and approximate values of $\overline{M}(\rho)$. The left, middle, and right graphs correspond to cases (A), (B), and (C), respectively. 
The values from the exact analysis are calculated by (\ref{eq-mean-full}) whereas those from the approximate analysis corresponding to cases (A), (B), and (C) are calculated by using (\ref{eq-M_Q-case-A})--(\ref{eq-M_R-case-A}), (\ref{eq-M_Q-case-B})--(\ref{eq-M_R-case-B}), and (\ref{eq-M_Q-case-C})--(\ref{eq-M_R-case-C-x}), respectively. 
We can see from the graphs that $\overline{M}(\rho)$ rapidly decreased as $\rho$ increased. In addition, when $N$ was larger, $\overline{M}(\rho)$ became smaller because the interference at the intersection became higher. We can also see that our approximation performed well except for region where $\rho > 0.8$ in case (A). %, which was larger than $0.8$ in case (A). 
Furthermore, the errors increased when $N$ was small. Similar to the evaluation of $p(r)$, this is because 
we assume that the queue length $N$ is sufficiently large. % in our approximation. 
%However, the optimal $\rho_{\ast}$ is usually around $0.05\sim0.2$ (see also Section~\ref{subsec-eval-opt}), and thus, we regard our approximate formulae for $\overline{M}(\rho)$ as being sufficiently accurate in realistic settings. 
%In addition, we can find that the errors in the graph in Figure~\ref{fig:exp-1-case-A} barely affected the approximation for $\overline{M}(\rho)$. 
%

\subsection{Effectiveness of Optimization Method}\label{subsec-eval-opt}

\begin{figure*}[!t]
\centering
\includegraphics[width=2.0in]{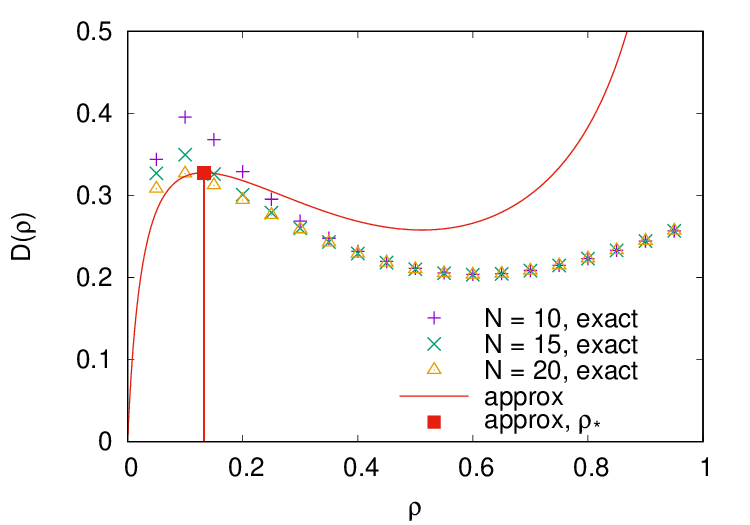}~~~
\includegraphics[width=2.0in]{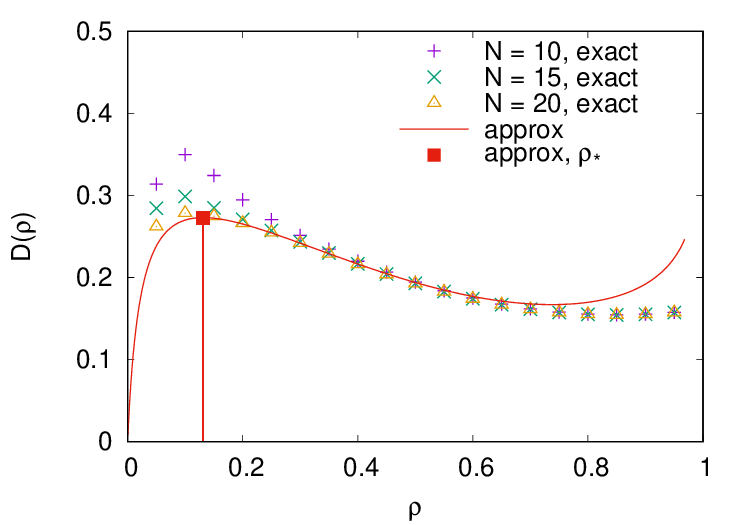}~~~
\includegraphics[width=2.0in]{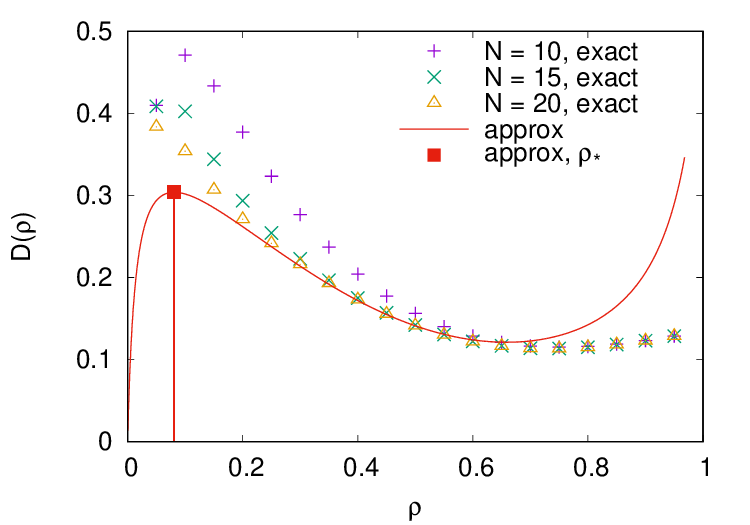}
\caption{Comparison of values of $D(\rho)$ from exact/approximate analysis with different $\rho$ and $N$.
Left, middle, and right graphs correspond to cases (A), (B), and (C). Vertical 
line represents $\rho_{\ast}$ that maximizes approximate $D(\rho)$. 
}
\label{fig:exp-3}
\vskip -7pt
\end{figure*}

We next provide the evaluation results for the broadcast rate optimization method. % presented in Section~\ref{subsec-optimize}. 
Figure~\ref{fig:exp-3} shows the results for the objective function $D(\rho)$ with different $\rho$ and $N$. We also plotted the optimal $\rho_{\ast}$'s that maximized the approximate $D(\rho)$ in the same graphs. 
The left, middle, and right graphs correspond to cases (A), (B), and (C), respectively. 
All values in the graphs were calculated by using (\ref{eq-mean-full}) or approximate formulae in Section~\ref{sec-app}. 
We first focus on the results from the exact analysis. We can see from the graphs that there are local maximum values in the domain $\rho \le 0.5$ in all cases. The figure shows that the optimal $\rho$'s achieved roughly 1.5 times higher $D(\rho)$ at the maximum than those when $\rho = 0.5$ (unnecessarily high case). 
We can see a similar tendency in all cases, however, 
the right graph shows that $D(\rho)$ in case (C) decreased more significantly
than the other cases as $\rho$ increased. 
Recall that neighbors of the transmitter in case (C) exist in $S_{R_x}$ and $S_Q$
while those in case (A) exist in $S_R{_x}$, $S_{R_y}$, and $S_Q$ and those in case (B) exist 
in $S_{R_x}$ and $S_Q$ of only the left  part of the transmitter. 
Thus, if $\rho$ increases, the interference in case (C) becomes higher than (B) 
and the number of potential receivers (i.e., vehicles not transmitting) in case (C) 
becomes less than in case (A). This is why our broadcast rate
optimization has more significant effect in case (C) than cases (A) and (B). 
%the broadcast optimization in case (C) is much larger than 
%
Furthermore, when $\rho$ approached $1$, $D(\rho)$ slightly increased. This is because we fixed $\rho_0$ of the vehicles in $S_R$. Thus, if $\rho$ increases, the transmitter has more of a chance to transmit to vehicles in $S_R$ even though $p(r)$ becomes smaller. 
However, $\rho > 0.5$ is unrealistic when considering a receiving time or other computational time. 
Thus, we consider $\rho_{\ast} < 0.5$.  %the domain $\rho \le 0.5$. 
%Indeed, in our settings, we can see that unique optimal $\rho_{\ast}$'s exist in that domain. 

We next discuss the accuracy of our approximation. We can see from the graphs that the approximated values of $D(\rho)$ fit well to those from the exact analysis in the domain $\rho \le 0.5$. We can also see that the value of $\rho_{\ast}$ was not very sensitive to the value of $N$ in all cases. This suggests that our approximate $D(\rho)$ not depending on $N$ is valid. However, the errors 
became larger due to the assumptions that $N$ is sufficiently large and $(1 - \rho)T > 1$. 
%between the values from the exact and approximate analysis 
 %when $N$ was smaller or $\rho$ was larger
%This is because in our approximation, we assume 
%Thus, in the region where $N$ is small or $\rho$ is close to $1$, the approximate formulae becomes less accurate. 
%In addition, the errors in the left graph in the domain $\rho \ge 0.5$ were larger than those in the middle and right ones. 
%The reason for this is that we approximated the Euclidean distance from the receiver in $S_{R_y}$ to the vehicles in $S_Q$ by the Manhattan distance, which is larger than the Euclidean one. Thus, the interference from the vehicles in $S_Q$ to the receivers in $S_{R_y}$ was underestimated. 
%Consequently, the approximate values of $D(\rho)$ were larger than those in the exact analysis. 
Fortunately, the errors are relatively small in the domain where $\rho_{\ast}$ existed and $\rho \le 0.5$. Therefore, we can say that the optimization with our approximation provides a good guideline for the optimal $\rho_{\ast}$.

\begin{figure*}[!t]
\centering
\includegraphics[width=2.0in]{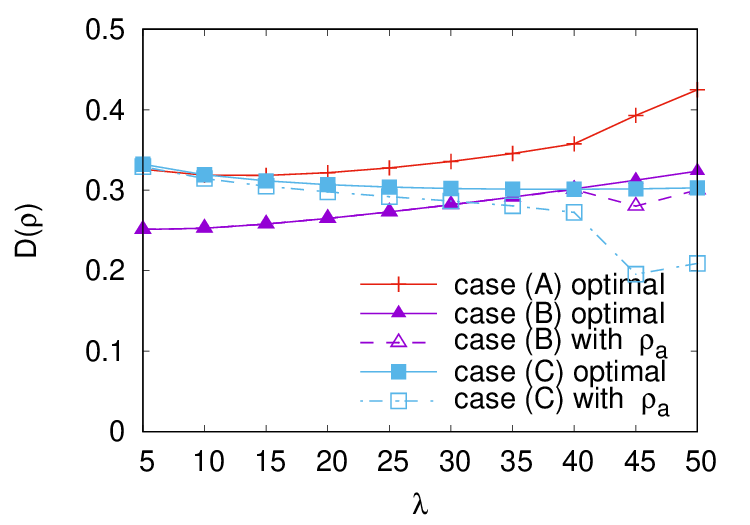}~~~
\includegraphics[width=2.0in]{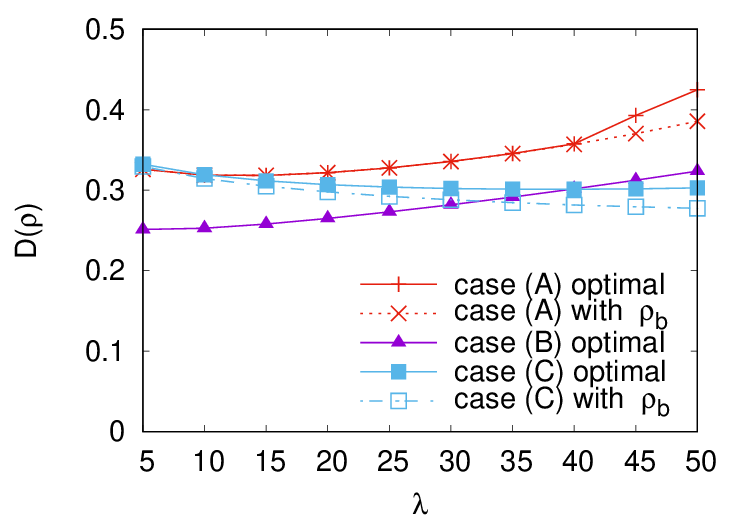}~~~
\includegraphics[width=2.0in]{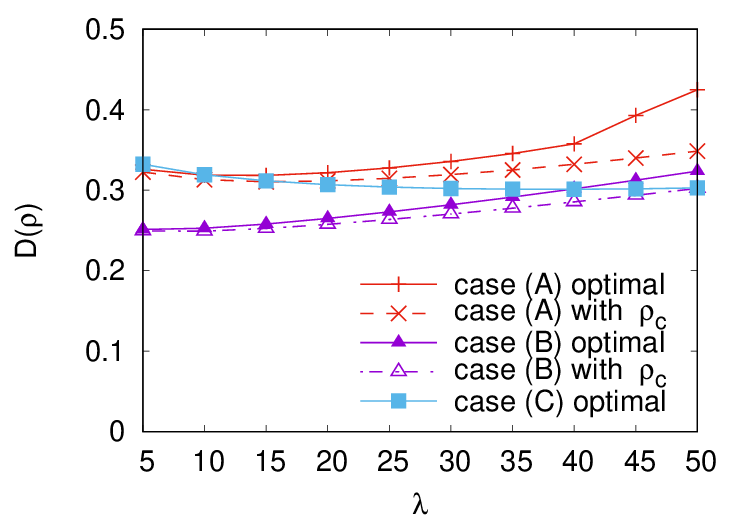}
\caption{
Comparison of %approximate $D(\rho_{\ast})$ in cases (A)--(C)
$D(\rho_a)$ (left), $D(\rho_b)$ (middle), and $D(\rho_c)$ (right) 
in cases (A)--(C) with different $\lambda$, 
where $\rho_a$, $\rho_b$, and $\rho_c$ are $\rho_\ast$ in cases (A)--(C), respectively
(dashed lines). 
Solid lines represent approximate $D(\rho_{\ast})$ in cases (A)--(C). 
%Comparison of values of $D(\rho)$ from exact/approximate analysis with different $\rho$ and $N$.
%Left, middle, and right graphs correspond to cases (A), (B), and (C). Vertical 
%line represents $\rho_{\ast}$ that maximizes approximate $D(\rho)$. 
}
\label{fig:exp-4}
\vskip -7pt
\end{figure*}

We now discuss relationship between $\rho_{\ast}$'s in cases (A)--(C). 
As we mentioned in Section~\ref{subsec-optimize}, 
our optimization problem depends on the position of a transmitter. 
Figure~\ref{fig:exp-4} compares results of the values of 
$D(\rho_a)$, $D(\rho_b)$, and $D(\rho_c)$ in cases (A)--(C) with different $\lambda$ and $N = 25$, 
where $\rho_a$, $\rho_b$ and $\rho_c$ are equal to $\rho_{\ast}$ in cases (A)--(C), respectively. 
We also plotted $D(\rho_\ast)$ in all three cases (A)--(C). 
% with different $\lambda$ and $N = 25$. 
%We also plotted $D(\rho_a)$, $D(\rho_b)$, and $D(\rho_c)$
%in cases (A)--(C) as the dashed and dotted lines, 
%In addition, we fixed $N = 25$. 
We can see from the figure that $D(\rho_\ast)$ 
and $D(\rho_a)$, $D(\rho_b)$, and $D(\rho_c)$  in cases (A)--(C)
took similar values when $\lambda$ was smaller than $45$. 
%This fact is supported by Fig.~\ref{fig:exp-3}, in which
%$D(\rho)$ was insensitive to $\rho$ around $\rho_\ast$. 
This fact suggests that $\rho_\ast$ is almost insensitive to 
cases (A)--(C) if $\lambda$ is not very high. Thus 
%Therefore, 
by adopting a $\rho \in \{\rho_a, \rho_b, \rho_c\}$ 
to determine the broadcast rate of all vehicles in $S_Q$, 
we can roughly maximize $D(\rho)$ regardless of the transmitter position. 
The reason why the difference between $D(\rho_a)$ and 
the optimal $D(\rho_\ast)$ in case (C) [left graph]
and that between $D(\rho_c)$ and the optimal $D(\rho_\ast)$ in case (A) [right graph]
increased as $\lambda$ increased is that if $\lambda$ is higher, the impacts of 
interferes and receivers in $S_{R_Y}$ becomes larger and thus the difference 
between $\rho_a$ and $\rho_c$ becomes larger. 
%in all the cases. 
% if the traffic intensity 
%in 
%the running segment is not high. 
%

\begin{figure}[!t]
\centering
\includegraphics[width=2.0in]{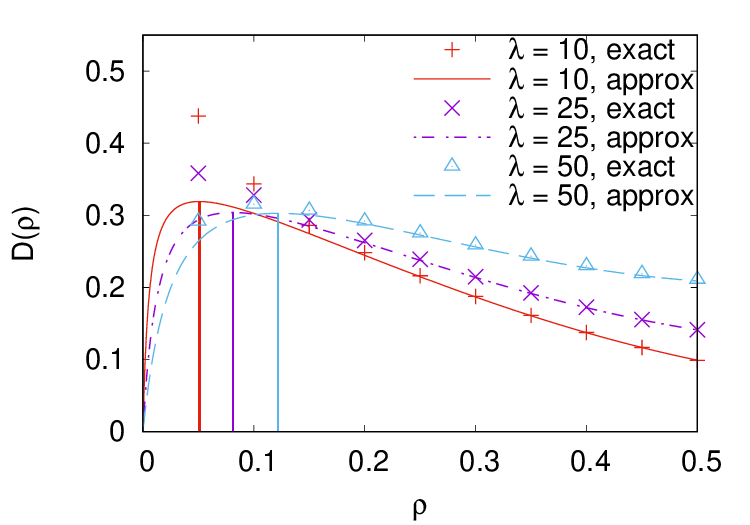}
\caption{
Comparison of $D(\rho)$ from exact/approximate analysis with different $\rho$ and $\lambda$
in case (C). Vertical line represents $\rho_{\ast}$ that maximizes approximate $D(\rho)$. 
}
\label{fig:exp-4-2}
\end{figure}

%
We also evaluate the impact of $\lambda$ on $\rho_\ast$. % in case (C), i.e., $\rho_c$. 
Figure~\ref{fig:exp-4-2} shows the results for $D(\rho)$ of
approximate and exact analysis in case (C) when varying $\lambda$.
From the figure, we can see that $\rho_\ast$ increased subject to $\lambda$. 
%The figure also shows that the approximate values for $\rho_{\ast}$ 
%again fitted well to those from the exact analysis. 
Recall here that $\lambda$ is the key parameter for determining the optimal
$\rho$ (see Section~\ref{subsec-optimize}). As a result, the results in 
Figures~\ref{fig:exp-4} and \ref{fig:exp-4-2} show that
we can determine the optimal broadcast rate of the vehicles in $S_Q$ by only observing the 
traffic intensity $\lambda$ because it is almost insensitive to cases (A)--(C) and the queue length. 

\subsection{Vehicles at Other Locations}\label{subsec-other-pos}

\begin{figure}[!t]
\begin{minipage}[t]{0.48\hsize}
\centering
\includegraphics[width=1.72in]{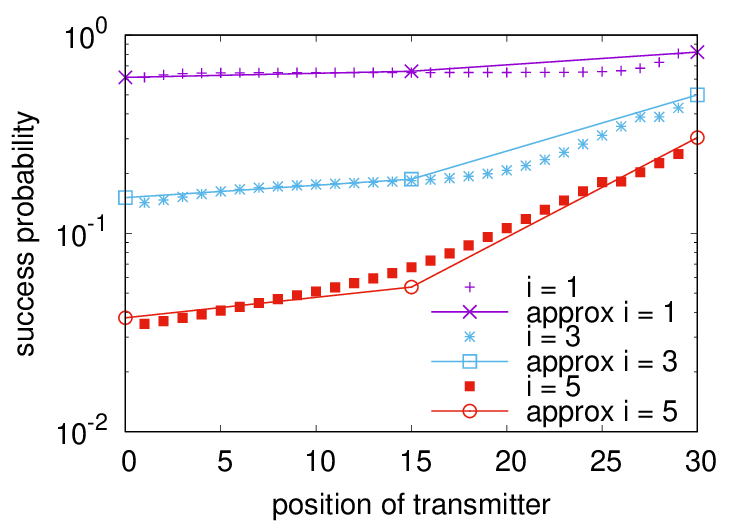}
\caption{Comparison of $p(il_v)$ with different position of transmitter at $d$-th vehicle from intersection. 
$N = 30$ and $\rho = 0.1$. Dashed line was plotted by interpolating the approximate formulae for cases (A)--(C).}
%$d = 0$, (B) $d = 30$, and (C) $d=15$. }
\label{fig:exp-7}
\end{minipage}
~
\begin{minipage}[t]{0.49\hsize}
\centering
\includegraphics[width=1.72in]{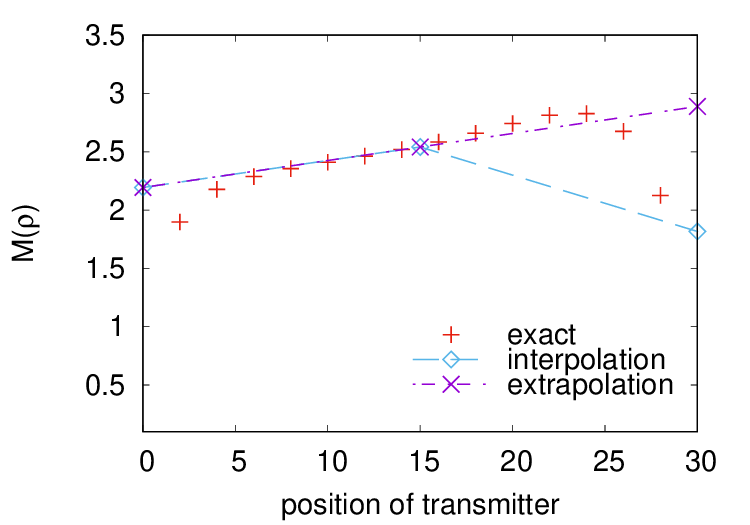}
\caption{Comparison of $\overline{M}_Q(\rho)$ when varying position of transmitter from intersection. $N = 30 $ and $\rho = 0.1$. Dashed line was plotted by interpolating approximation for cases (A) $d = 0$, (B) $d = 30$, and (C) $d=15$. Dashed and dotted line was plotted by extrapolating those for cases (A) and (C). }
\label{fig:exp-7-sum}
\end{minipage}
\vskip -7pt
\end{figure}

We next consider the case where a transmitter is at other locations than cases (A)--(C)
, i.e., not at the intersection, at the end of the queue, or the middle of the queue. 
Since our approximation assumes that $N$ is sufficiently large and only considers special cases (A)--(C), 
%Rdoes not consider the distance from the end of the queue, 
we cannot obtain closed-form formulae for $p(r)$ or $\overline{M}(\rho)$ in general cases. 
%Thus, we aim to roughly estimate their performance by interpolating the values of approximate formulae for cases (A)--(C). 
However, vehicles at other locations in the queue can be considered as being in an intermediate state between the vehicle at the intersection and that at the end of the queue. Thus, it is expected that we can roughly estimate their performance by interpolating the values of approximate formulae for cases (A)--(C). 
Figure~\ref{fig:exp-7} shows the values of $p(i l_v)$ when varying the positions of the transmitter and the distance to the receiver $i$ in $S_Q$. We fixed $N = 30$ and $\rho = 0.1$, and the $y$-axis was in log scale. In the graph, the dashed lines represent the interpolation line using the approximation formulae for cases (A)--(C). %, where cases (A), (B), and (C) correspond to $d = 0$, $d = N/2 = 15$, and $d =30$. 
%In other words, the dashed line shows the {\it log-linear interpolation}. 
From the figure, we can see that the interpolation can roughly estimate the values of $p(i l_v)$ in all cases. We can also see that when $i$ increased, the results from the exact analysis became close to log-linear, whereas when $i$ was small, they tended to be a constant value. This is because if the receiver is closer to the transmitter, the impacts of the intersection or the end of the queue rapidly disappear as the distance from the transmitter to them increases. 
Similarly, Figure~\ref{fig:exp-7-sum} shows the results for $\overline{M}(\rho)$, i.e., the mean number of successful receivers in $S_Q$ when varying the positions of the transmitter $d$. Note that the $y$-axis is in linear scale. The dashed line was plotted by interpolating the results of the approximation for the three cases. The dashed and dotted line was plotted by extrapolating the approximation for cases (A) and (C). We can see that if the positions of the transmitter were close to the middle of the queue, i.e., $N / 2$, the extrapolation well estimated the exact values. However, if the transmitter was close to the intersection or the end of the queue, the error increased. This tendency is similar to that in Figure~\ref{fig:exp-7}. 
As a result, we can conclude the following. If the transmitter is close to the middle of the queue, we can use extrapolation on the basis of the approximate formulae for cases (A) and (C); otherwise, the approximate formulae for cases (A) and (B) should be used instead. 

%
%\begin{figure}[!t]
%\centering
%\begin{minipage}[t]{0.48\hsize}
%\centering
%%\includegraphics[width=2.3in]{imgs/exp_4_lam_2.eps}
%%\includegraphics[width=1.72in]{imgs/exp_4_lam_2.eps}
%\includegraphics[width=1.72in]{exp_4_lam_2.eps}
%%\includegraphics[width=1.72in]{exp_4_lam_2.eps}
%\caption{
%Comparison of $D(\rho)$ from exact/approximate analysis with different $\rho$ and $\lambda$
%in case (C). Vertical line represents $\rho_{\ast}$ that maximizes approximate $D(\rho)$. 
%}
%\label{fig:exp-4-2}
%\end{minipage}
%~
%\begin{minipage}[t]{0.48\hsize}
%\centering
%\includegraphics[width=1.72in]{exp_7.eps}
%%\includegraphics[width=1.72in]{imgs/exp_7.eps}
%\caption{Comparison of $p(il_v)$ with different position of transmitter from intersection. %$d$-th vehicle ). 
%Dashed line was plotted by interpolating approximate formulae for cases (A)--(C).}
%%$d = 0$, (B) $d = 30$, and (C) $d=15$. }
%\label{fig:exp-7}
%\end{minipage}
%\vskip -7pt
%\end{figure}
%

\section{Conclusion}\label{sec-conclude}

In this paper, we proposed an optimization method for the broadcast rate in 
vehicle-to-vehicle (V2V) communications at an intersection based on
theoretical analysis. 
%theoretically analyzed the performance of vehicle-to-vehicle (V2V) broadcast communications at an intersection. 
Since the theoretical values of the probability of successful transmission and the mean number of successful receivers are non-analytical, 
we provided closed-form approximations for them. 
%We also provided a method for optimizing the broadcast rate of vehicles in the queuing segment. 
By using the closed-form formulae, we can obtain the optimal broadcast rate without time-consuming numerical computation. 
Through numerical examples, we found that our 
broadcast rate optimization achieved roughly 1.5 times higher performance than
the case without broadcast rate control. 

To maintain mathematical tractability, we assumed a simple channel 
model and media access control (MAC) layer in this paper, e.g., Rayleigh fading assumption or ALOHA. 
Thus, the generalization of the distribution of vehicles and fading 
are for future work. In addition, power control can be a good 
solution for the interference problem at an intersection. Therefore, 
the joint modeling and optimization of the broadcast rate and the
transmission power of vehicles are also for future work. 
In addition, in our optimization method, we assume that the 
traffic intensities in running segments are given. However, 
vehicles/road side units need to infer these values in a practical situation,
e.g., by measuring the distance to the vehicle running at the front. 
Such inference schemes and their impacts on our optimization method
are also for future work.

\appendices

%% you can choose not to have a title for an appendix
%% if you want by leaving the argument blank
%\section{}

\section{Approximation Methodology}\label{appen-approx}
In this appendix, we give detailed explanations 
for the derivation of the approximate formulae presented in 
Section~\ref{sec-app}. 
For later use, we first introduce approximate formulae for
$q_{n_0, T} (r \mid n_1)$ defined as
\begin{equation}
q_{n_0,T}(r \mid n_1)  = \sum_{m=1}^{n_1}\log\left( 1 + T\left({r  \over n_0 + m}\right)^{\alpha}\right),
\label{eq-def-q_i(T)}
\end{equation}
which plays an important role in our approximation method for $\calL_{I_Q}( T r^{\alpha} \mid d)$. 
%(for details, see Appendix~\ref{appen-auxiliary}). 
Derivation of the following formulae with several auxiliary results
are given in Appendix~\ref{appen-auxiliary}. 

\medskip
\noindent
{\bf Approximate formulae of $q_{n_0, T}(r \mid n_1)$:} 
$q_{n_0, T}(r \mid n_1)$ can be approximated as follows.
\begin{enumerate}\renewcommand{\labelenumi}{(\roman{enumi})}
\item If $r < T^{-{1\over\alpha}}(n_0 + 1)$, 
\begin{eqnarray}
\lefteqn{
q_{n_0, T}(r \mid n_1) 
} &&
\nonumber
\\
&\approx&
\left\{
\begin{array}{ll}
\zeta(\alpha)T r^{\alpha},  & n_0 = 0, \\
{T \over \alpha - 1}\left[ {1 \over n_0^{\alpha - 1}} - {1 \over (n_0 + n_1)^{\alpha - 1}}\right]r^{\alpha},
   & n_0 > 0,
\end{array}
\right.
\nonumber
\end{eqnarray}
where $\zeta(\alpha)$ $(\alpha > 0)$ denotes the Riemann zeta function defined as
\begin{equation}
\zeta(\alpha) =  \sum_{k=1}^{\infty}{1 \over k^{\alpha}}.
\label{eq-def-zeta}
\end{equation}
\item If $T^{- {1 \over \alpha}} (n_0 + 1) \le r < T^{- {1 \over \alpha}}(n_0 + n_1)$, 
\begin{eqnarray}
\lefteqn{
q_{n_0, T}(r \mid n_1)
\approx
(\alpha + \kappa_{1, \alpha} - \kappa_{2, \alpha}) T^{1 \over \alpha}r
} &&
\nonumber
\\
&-&
 \alpha \left(n_0 + {1 \over 2} \right) \log T^{1 \over \alpha} r
- {T r^{\alpha} \over (\alpha - 1)(n_0 + n_1)^{\alpha - 1}}
\nonumber
\\
&-&
{1\over 2}\log\left(1 + T\left({r \over n_0 + n_1}\right)^{\alpha}\right)
- \underline{\psi}_{n_0, T}(r) 
\nonumber
\\
&&
\nonumber
\\
&-&
{1 \over 2}  \log \left(1 + {1 \over T}\left({n_0 \over r}\right)^{\alpha}\right)
+
\kappa_{1, \alpha } 
\nonumber
\\
&+&
\alpha \log {n_0 ! \over \sqrt{2 \pi}} + \log 2,\quad
\label{approx-fand}
\end{eqnarray}
where
\begin{equation}
\underline{\psi}_{n_0, T}(r) = 
n_0 \sum_{k=1}^{\infty}
{(-1)^{k+1}  \over k (\alpha k + 1) T^k}
\left(
{ n_0
\over
r
}
\right)^{\alpha k}.\quad
\label{eq-def-under-psi}
\end{equation}
\item If $T^{- {1 \over \alpha}}(n_0 + n_1) \le r$, 
\begin{eqnarray}
%\lefteqn{
q_{n_0, T}(r \mid n_1) 
&\approx &
\alpha n_1 \log T^{1\over \alpha}r 
%}\quad\qquad&&
\nonumber
\\
&{}& 
+
{1 \over T} \left[ {n_0 + n_1 \over \alpha + 1} + {1 \over 2} \right] \left({n_0 + n_1 \over r}\right)^{\alpha}
\nonumber
\\
&{}&
-
{1 \over T} \left[ {n_0  \over \alpha + 1} + {1 \over 2} \right] \left({n_0 \over r}\right)^{\alpha}
\nonumber
\\
&&
-
\alpha \left(n_1 + n_0 + {1 \over 2}\right)
\log(n_0 + n_1)
\nonumber
\\
&{}&
+ \alpha (n_0 + n_1)
+ \alpha \log {n_0! \over \sqrt{2 \pi}  }.\quad
\label{approx-fand-3}
\end{eqnarray}
\end{enumerate}
%

%\subsection{Derivation of approximate formulae in case (A)}
\subsection{Derivation of Equations~(\ref{eq-prob-app-case-A})--(\ref{eq-prob-app-case-A-r})}
%\subsection{Derivation of Results in Section~\ref{subsec-p(r)-case-A}}
\label{appen-subsec-p(r)-case-A}

We first prove that (\ref{eq-prob-app-case-A}) is true. To do this,
we temporarily assume that $n_+ = n_- = N$. This assumption 
will be removed later by considering a sufficiently large $N$. 
Since a receiver is the $i$-th vehicle from the intersection, 
by substituting $s =  T (i l_v)^{\alpha}$ into (\ref{eq-lap-I_Q}), 
we have
\begin{align}
&\calL_{I_Q}(T (il_v)^{\alpha} \mid i l_v) 
= \prod_{m =-n_1, m \neq i,0}^{n_+}\left[
{|i - m|^{\alpha} \rho \over |i - m|^{\alpha} + T  i^{\alpha}}
+ 1 - \rho
\right]
\nonumber
\\
&= 
%{1 + T \over 1 + (1 - \rho)T} 
K(\rho) \prod_{m=1, m \neq i}^{2N}
{1
+ (1 - \rho) T \left|{ i\over i - m }\right|^{\alpha}
\over 1 + T \left|{ i \over i - m }\right|^{\alpha}}
\nonumber
\\
&=
%{1 + T \over 1 + (1 - \rho)T}
K(\rho) \prod_{m=1}^{N- i}
{1
+ (1 - \rho) T \left({ i \over m }\right)^{\alpha}
\over 1 + T \left({ i \over m }\right)^{\alpha}}
%\nonumber
%\\
%&{}&
%\times
\prod_{m=1}^{N + i}
{1
+ (1 - \rho) T \left({ i \over m }\right)^{\alpha}
\over 1 + T \left({ i \over m }\right)^{\alpha}},
%\nonumber
%\\
\label{eq-lap-I_Q-2}
\end{align}
where $K(\rho)$ is given in (\ref{eq-def-K(rho)}). 
%We now define a function $q_{n_0, T} (r \mid n_1)$ as 
It follows from  (\ref{eq-def-q_i(T)}) and  (\ref{eq-lap-I_Q-2}) that 
\begin{align}
&
\log \calL_{I_Q}(T  (il_v)^{\alpha} \mid i l_v)
=  \log K(\rho) + 
q_{0, (1-\rho) T}(i \mid N + i) 
\nonumber
\\
&
+ q_{0, (1-\rho) T}( i \mid N - i)
 - q_{0, T}( i \mid N + i) - q_{0, T}( i \mid N - i).
\label{eq-lap-d_0=0-1002}
\end{align}
We now assume that $N$ is sufficiently large. 
Recall here that $(1 - \rho) T \ge 1$ and thus $((1 - \rho)T)^{- {1 \over \alpha}} < 1$. 
Therefore, we can apply (\ref{approx-fand}) % in Appendix~\ref{appen-auxiliary}
%Proposition~\ref{prop-approx-fand}~(ii) 
and thus obtain
\begin{equation}
q_{0, (1 - \rho) T}(i \mid \infty) - q_{0, T}(i \mid \infty)
\approx
\xi_{\alpha, T}(\rho) i -{1\over 2}
\log (1 - \rho),
%\nonumber
%\\
\label{eq-app-q-i-1}
\end{equation}
where $\xi_{\alpha, T}(\rho)$ is defined in (\ref{eq-def-xi}). 
%}
It then follows from (\ref{eq-lap-d_0=0-1002}) that
\begin{equation}
\log \calL_{I_Q}( T  (il_v)^{\alpha} \mid il_v)
\approx
\log {K(\rho) \over 1 - \rho}
%\log{1 + T \over (1 - \rho)(1 + (1 - \rho)T)}
+ 
2 \xi_{\alpha, T}(\rho) i.\qquad
\label{eq-lap-add-1}
\end{equation}
In addition,
by substituting $d = i l_v$ and $r = i l_v$ into 
(\ref{eq-lap^X_I_R}) and (\ref{eq-lap^Y_I_R}), 
we can easily obtain
\begin{align}
\log\calL_{I_R^X}( T (i l_v)^{\alpha}) &=
-  \rho_0 \lambda_x C_{\alpha, T}^X i l_v,
\label{eq-cal_I_X-C_T}
\\
\log\calL_{I_R^Y}(T (i l_v)^{\alpha} \mid i l_v) 
&=
- \rho_0 \lambda_y C^Y_{\alpha,T} i l_v.
\label{eq-cal_I_Y-C_T}
\end{align}
%%
%\begin{eqnarray}
%\log\calL_{I_R^X}( T (i l_v)^{\alpha}) 
%&=&
%- 2 \rho \lambda_x  \int_0^{\infty} {1 \over (x/T^{1\over \alpha}il_v)^{\alpha} + 1} \rmd x
%\nonumber
%\\
%&=&-  \rho \lambda_x C_{\alpha, T}^X i l_v.
%\label{eq-cal_I_X-C_T}
%\end{eqnarray}
%%
%where $C_{\alpha, T}^X$ is given in (\ref{eq-def-C_T}). 
%Similarly, from (\ref{eq-lap^Y_I_R}), we obtain
%%
%
%\begin{eqnarray}
%\lefteqn{\log\calL_{I_R^Y}(T (i l_v)^{\alpha} \mid i l_v) }
%\qquad
%&&
%\nonumber
%\\
%%&=&
%%- 2 \rho_0 \lambda_y \int_0^{\infty} {T (il_v)^{\alpha} \over (y^2 + (il_v)^2)^{\alpha \over 2} + T (il_v)^{\alpha}} \rmd y
%%\nonumber
%%\\
%&=&
%- 2 \rho_0 \lambda_y \int_{0}^\infty {T \over ((y / il_v)^2 + 1)^{\alpha \over 2} + T} \rmd y
%\nonumber
%\\
%&=&
%- \pi \rho_0 \lambda_y C^Y_{\alpha,T} i l_v.
%\label{eq-cal_I_Y-C_T}
%\end{eqnarray}
where  $C_{\alpha, T}^X$ and 
$C_{\alpha, T}^Y$ are given in 
(\ref{eq-def-C_T}). 
%(\ref{eq-lap-Y-alpha-4}) and (\ref{eq-lap-Y-alpha-2}), 
%leads to  (\ref{eq-def-C_T}) and
%%
%\begin{eqnarray}
%\log\calL_{I_R^X}(T (i l_v)^{\alpha}) 
%&=&
%- \pi \rho_0 \lambda_x C^X_{\alpha,T} i l_v,
%\label{eq-cal_I_X-C_T}
%\\
%%\label{eq-cal_I_Y-C_T}
%\log\calL_{I_R^Y}(T (i l_v)^{\alpha} \mid i l_v) 
%&=&
%- \pi \rho_0 \lambda_y C^Y_{\alpha,T} i l_v.
%\label{eq-cal_I_Y-C_T}
%\end{eqnarray}
%
As a result, combining (\ref{eq-lap-add-1})--(\ref{eq-cal_I_Y-C_T}) with 
(\ref{eq-cov-full}) yields (\ref{eq-prob-app-case-A}). 

We now move on to the derivation of (\ref{eq-prob-app-case-A-r-x}). 
To proceed, we assume that $r = r_0 l_v$ $(r_0 \in \bbN$). 
This assumption will be removed later by 
extending the result to an arbitrary $r \in \bbR$. 
By following the same arguments in the derivation of 
(\ref{eq-lap-I_Q-2}) and (\ref{eq-lap-d_0=0-1002}), we have
\begin{eqnarray}
\lefteqn{
\log \calL_{I_Q}(T r^{\alpha} \mid r) 
= \underset{m \neq 0}{\sum_{m =-n_1}^{n_+}}\log \left[
{|r_0 - m|^{\alpha} \rho \over |r_0 - m|^{\alpha} + T  i^{\alpha}}
+ 1 - \rho
\right]
} &&
\nonumber
\\
&=&
%\log (1 - \rho) 
%+ \log {1 + T \over 1 + (1 - \rho)T}
q_{0, (1-\rho) T}(r_0 \mid N + r_0) 
+
q_{0, (1-\rho) T}( r_0 \mid N - r_0)
\nonumber
\\
&-&
q_{0, T}( r_0 \mid N + r_0) 
%\nonumber
%\\
%&{}&
- q_{0, T}( r_0 \mid N - r_0)
+ \log (1 - \rho) K(\rho).
\nonumber
\end{eqnarray}
Thus, similar to (\ref{eq-app-q-i-1}) and (\ref{eq-lap-add-1}), 
letting $N \to \infty$ and $r_0 = r / l_v \in \bbR$ and 
applying  (\ref{approx-fand}) %in Appendix~\ref{appen-auxiliary}
leads to 
\begin{eqnarray}
\log \calL_{I_Q}(T r^{\alpha} \mid r) 
\approx 
2 \xi_{\alpha, T}(\rho){r \over l_v}
+
\log K(\rho). 
%\log {1 + T \over 1 + (1 - \rho)T}.
\nonumber
\end{eqnarray}
From this, (\ref{eq-cov-full}), (\ref{eq-cal_I_X-C_T}), and (\ref{eq-cal_I_Y-C_T}),
we obtain (\ref{eq-prob-app-case-A-r-x}). 

Finally, we derive (\ref{eq-prob-app-case-A-r}). 
Since the receiver is in  $S_{R_y}$ at distance $r$
from the intersection, its Euclidean distance from the $i$-th vehicle from 
the intersection is equal to $\sqrt{r^2 + (il_v)^2}$. However, applying this
Euclidean distance to $\calL_{I}(T r^{\alpha})$
leads to mathematically intractable analysis. Thus, we approximate this 
by using Manhattan distance, which is equal to $r + il_v$. 
Similar to the previous case, we assume that $r = r_0 l_v$ $(r_0 \in \bbN)$. 
Then, the Laplace transform of
$I_Q$ can be approximated as
%It then follows from (\ref{eq-lap-I_Q}) that
%
\begin{eqnarray}
%\lefteqn{
\calL_{I_Q}( T r^{\alpha} \mid r ) 
&=&
\prod_{m=1}^{n_-}
{1 + (1 - \rho) T \left({r_0 \over r_0 + m}\right)^{\alpha}
\over
1 + T \left({r_0 \over  r_0 + m}\right)^{\alpha}
}
%} \qquad \qquad&&
\nonumber
\\
&{}&
\times
\prod_{m=1}^{n_+}
{1 + (1 - \rho) T \left({r_0 \over r_0 + m}\right)^{\alpha}
\over
1 + T \left({r_0 \over  r_0 + m}\right)^{\alpha}
}.\qquad
\label{eq-L_Q_ii-def-2}
\end{eqnarray}
%}
%
By considering sufficiently large $n_+$ and $n_-$, we obtain
\begin{eqnarray}
\log \calL_{I_Q}(T r^{\alpha} \mid r) 
&=&
2 \left(q_{r_0, (1- \rho)T}(r_0 \mid \infty) -q_{r_0, T}(r_0 \mid \infty)\right).
\nonumber
\end{eqnarray}
Since $((1 - \rho)T)^{- {1 \over \alpha}} < 1$, 
we can apply (\ref{approx-fand}) %in Appendix~\ref{appen-auxiliary} 
%Proposition~\ref{prop-approx-fand} (ii) 
to this and thus obtain
%
%{\small
\begin{eqnarray}
\lefteqn{
\log \calL_{I_Q}( T r^{\alpha}  \mid r) 
\approx
2 \xi_{\alpha, T}(\rho) r_0
- \left(r_0 + {1 \over 2}\right) \log (1 - \rho)
} &&
\nonumber
\\
&{}&
- \log {1 + {1 \over (1 - \rho) T} \over 1 + {1 \over T}}
- 2(\underline{\psi}_{r_0, (1-\rho)T}(r_0) - \underline{\psi}_{r_0, T}(r_0))
\nonumber
\\
&\approx&
2 \left[
\xi_{\alpha, T}(\rho) - \log (1 - \rho)
- {\rho \over (\alpha + 1)(1- \rho)T}
\right] r_0
\nonumber
\\
&{}&
-
\log (1 - \rho)
+ \log{ (1 - \rho)(1 + T) \over 1 + (1 - \rho)T },\quad
\label{eq-add-app-Y-tmp}
\end{eqnarray}
%}
%
where we use the following approximation [see (\ref{eq-def-under-psi})] % in Appendix~\ref{appen-auxiliary} of \cite{Kimu17}]
\begin{eqnarray}
\lefteqn{
\underline{\psi}_{r_0, (1 - \rho)T}(r_0) - \underline{\psi}_{r_0, T}(r_0) 
= r_0 \sum_{k=1}^{\infty}{(-1)^{k+1} \over k (\alpha k + 1)T^k} 
}\qquad\qquad &&
\nonumber
\\
&=&
{r_0 \over (\alpha + 1) T}{\rho \over 1 - \rho}
+ O(((1 - \rho)T)^2).
\end{eqnarray}
Although the receiver is on the $y$-axis, $p(r)$ can be 
calculated very similarly to Proposition~\ref{prop-cov}. 
Indeed, we obtain
\[
p(r) \approx
\calL_{I_Q}( T r^{\alpha} \mid r)
\calL_{I^X_R}( T r^{\alpha} ) \calL_{I^Y_R}(T r^{\alpha}  \mid r).
\]
As a result, substituting (\ref{eq-cal_I_X-C_T})--(\ref{eq-add-app-Y-tmp}) 
%and Proposition~\ref{prop-cov} 
into the above yields (\ref{eq-prob-app-case-A-r}). 
\qed
%

%\subsection{Derivation of approximate formulae in the case (B)}
%\subsection{Derivation of Results in Section~\ref{subsec-p(r)-case-B}}
\subsection{Derivation of Equations~(\ref{eq-app-case-B-i})--(\ref{eq-app-case-B-r})}
\label{appen-subsec-p(r)-case-B}
We begin with (\ref{eq-app-case-B-i}). Similar to (\ref{eq-lap-I_Q-2}), 
we assume that $n_+ = n_- = N$ and consider a sufficiently large $N$. 
Recall that the receiver is $i$-th vehicle from the end of the queue
and thus its distance from the intersection is equal to $(N - i)l_v$. 
Thus, by substituting $s =  T r^{\alpha}$ into (\ref{eq-lap-I_Q}) and
letting $r = i l_v$, we obtain
\begin{eqnarray}
\lefteqn{
\calL_{I_Q}(T (i l_v)^{\alpha} \mid (N - i)l_v) 
} &&
\nonumber
\\
&=&
\prod_{m=1}^{i-1}
{1
+ (1 - \rho) T \left({ i \over m }\right)^{\alpha}
\over 1 + T \left({ i \over m }\right)^{\alpha}}
\prod_{m=1}^{2N-i}
{1
+ (1 - \rho) T \left({ i \over m }\right)^{\alpha}
\over 1 + T \left({ i \over m }\right)^{\alpha}}
\nonumber
\\
&=&
K(\rho)
%{1 + T \over 1 + (1 - \rho) T}
\prod_{m=1}^{i}
{1
+ (1 - \rho) T \left({ i \over m }\right)^{\alpha}
\over 1 + T \left({ i \over m }\right)^{\alpha}}
%\nonumber
%\\
%&{}&
%\times
\prod_{m=1}^{2N-i}
{1
+ (1 - \rho) T \left({ i \over m }\right)^{\alpha}
\over 1 + T \left({ i \over m }\right)^{\alpha}}.
\nonumber
\end{eqnarray}
Using (\ref{eq-def-q_i(T)}), the above equation can be rewritten as 
follows.
\begin{eqnarray}
\lefteqn{
\log \calL_{I_Q}( T r^{\alpha} \mid  (N - i)l_v) 
= 
\log K(\rho)
%{%1 + T \over 1 + (1 - \rho) T}
+ q_{0, (1-\rho) T}(i \mid i) 
} &&
\nonumber
\\
&{}& 
+ q_{0, (1-\rho) T}( i \mid 2 N - i)
- q_{0, T}( i \mid i) - q_{0, T}( i \mid 2 N - i).\qquad
\label{eq-lap-d_0=0}
\end{eqnarray}
Since $((1 - \rho)T)^{- {1 \over \alpha}} < 1$, we can apply (\ref{approx-fand-3})
% in Appendix~\ref{appen-auxiliary} 
 and thus obtain
\begin{eqnarray}
\lefteqn{
q_{0, (1 - \rho) T}(i \mid i) - q_{0, T}(i \mid i)
}&&
\nonumber
\\
&\approx&
\left[ \log (1 - \rho) + {\rho \over T (1- \rho)(\alpha + 1) }
\right] i + {\rho \over 2 T (1- \rho)}.\quad
\label{eq-approx-add-case-2-i}
\end{eqnarray}
Furthermore, if we assume that $N$ is sufficiently large, 
we can use the approximation in (\ref{eq-app-q-i-1}). % in the proof of Theorem~\ref{thm-case-A}. 
As a result, substituting this and (\ref{eq-approx-add-case-2-i}) into 
(\ref{eq-lap-d_0=0}) and combining it with (\ref{eq-cal_I_X-C_T}) 
and (\ref{eq-app-p(r)-Lap_X-Q}) lead to  (\ref{eq-app-case-B-i}) and (\ref{eq-def-beta}). 

We now prove that (\ref{eq-app-case-B-i-x}) is true. Similar to the derivation of 
(\ref{eq-prob-app-case-A-r-x}) and (\ref{eq-prob-app-case-A-r}),
%the proof of the statements (ii) and (iii)
%of Theorem~\ref{thm-case-A}, 
suppose that $r =r_0 l_v$ $(r_0 \in \bbN)$. Then, the distance from the intersection
to the receiver is expressed as $(N - r_0)l_v$. 
Note here that if the $r_0$-th vehicle in  $S_Q$ from the transmitter is currently
transmitting, the transmission to the receiver fails.  Therefore, 
from  (\ref{eq-lap-I_Q}), (\ref{eq-def-q_i(T)}) and (\ref{eq-lap-d_0=0}), we obtain
\begin{eqnarray}
\lefteqn{
\log \calL_{I_Q}(T r^{\alpha} \mid (N - r_0)l_v)
= 
\log (1 - \rho) K(\rho)
%+ \log{1 + T \over 1 + (1 - \rho) T}
}\qquad\qquad &&
\nonumber
\\
&{}&
+
q_{0, (1-\rho) T}(r_0 \mid r_0)
+ q_{0, (1-\rho) T}( r_0 \mid 2 N - r_0)
\nonumber
\\
&{}&
- q_{0, T}( r_0 \mid r_0)
 - q_{0, T}( r_0 \mid 2 N - r_0).
%\label{eq-lap-d_0=0}
\nonumber
\end{eqnarray}
Plugging (\ref{eq-approx-add-case-2-i}) into the above and combining it with
(\ref{eq-cal_I_X-C_T}) and (\ref{eq-app-p(r)-Lap_X-Q}) yields (\ref{eq-app-case-B-i-x}). 

Finally, we consider (\ref{eq-app-case-B-r}). %statement (iii). 
In this case, if we assume that $r =r_0 l_v$,  (\ref{eq-lap-I_Q}) leads to 
\begin{eqnarray}
\calL_{I_Q}( T r^{\alpha} \mid (n_+ + r_0)l_v ) &=&
\prod_{m=1}^{n_- + n_+}
{1 + (1 - \rho) T \left({r_0 \over r_0 + m}\right)^{\alpha}
\over
1 + T \left({r_0 \over  r_0 + m}\right)^{\alpha}
}.
\nonumber
\end{eqnarray}
Therefore, by following the same arguments in the derivation of (\ref{eq-prob-app-case-A-r})
%in the proof of the
%statement (iii) of Theorem~\ref{thm-case-A} 
%(see e.g., (\ref{eq-L_Q_ii-def-2}) 
and (\ref{eq-add-app-Y-tmp}), 
we can readily show that  (\ref{eq-app-case-B-r}) holds. 
\qed

\section{Auxiliary Results}\label{appen-auxiliary}
%
%As shown in 
%the derivations of approximate formulae in Section~\ref{sec-app}, 
%the key idea of the approximation 
%is based on a rough estimation of the function $q_{n_0, T}(r \mid n_1)$
%defined in (\ref{eq-def-q_i(T)}). 
In this appendix, we 
discuss the approximation method for $q_{n_0, T}(r \mid n_1)$ defined in (\ref{eq-def-q_i(T)}). 
We first provide several lemmas, which are required 
for the approximation. 
All proofs of the lemmas are given in Appendix~\ref{appen-proof}. % of \cite{Kimu17}. 
Using these results, we derive approximate formulae of $q_{n_0, T}(r \mid n_1)$. 

To begin with, we define $\eta_{r,T}$ such that
\begin{equation}
\eta_{r,T} = \min\left\{m \in \bbN;  \left({r  \over n_0 + m }\right)^{\alpha} < {1 \over T}\right\} - 1.
\label{eq-def-over-i-T}
\end{equation}
In accordance with the value of $\eta_{r,T}$, we can consider three subcases:
(i) $\eta_{r,T} = 0$; (ii) $1 \le \eta_{r,T} \le n_1$; and (iii) $n_1 < \eta_{r,T}$. 
In what follows, we provide lemmas corresponding to each subcase. 
Note that we only use the results corresponding to  cases (ii) and (iii)
in the main part of our paper, 
but, we also consider case (i) $\eta_{r,T} = 0$
for completeness in this appendix.

We start with case (i) $\eta_{r,T} = 0$. By definition, this case suggests that
\begin{equation}
T \left({r \over n_0 + 1}\right)^{\alpha} < 1. 
\label{eq-case-iii-cond}
\end{equation}
We then have the following lemma, which provides an estimation of
$q_{n_0, T}(r \mid n_1)$ under the condition in which (\ref{eq-case-iii-cond}) holds. 
\begin{lem}\label{lem-case-1}
Suppose that $\eta_{r,T}  = 0$, i.e., (\ref{eq-case-iii-cond}) holds. 
If $n_0 = 0$, then 
\begin{equation}
q_{0, T}(r \mid n_1)  = 
 \zeta(\alpha) T r^{\alpha}  + O((Tr^{\alpha})^2) + O(T r^{\alpha} n_1^{-\alpha+1}),\quad
\label{eq-approx-q_i-case-1}
\end{equation}
%
%where $\zeta(\alpha)$ $(\alpha > 0)$ denotes the Riemann zeta function defined as
%%
%\begin{equation}
%\zeta(\alpha) =  \sum_{k=1}^{\infty}{1 \over k^{\alpha}},
%\label{eq-def-zeta}
%\end{equation}
%
otherwise, if $n_0 \ge 1$, then
\begin{eqnarray}
\lefteqn{
q_{n_0, T}(r \mid n_1)
=
T\left[{ n_0 \over \alpha - 1} + {1 \over 2}\right]
\left({r \over n_0}\right)^{\alpha}  
} \quad&&
\nonumber
\\
&-&
T \left[{ (n_0 + n_1)\over \alpha - 1} + {1 \over 2}\right]
\left({r \over n_0 + n_1}\right)^{\alpha}  
\nonumber
\\
&+&
O(T r^\alpha n_0^{-\alpha - 1}) + 
O(T r^\alpha (n_0 + n_1)^{-\alpha - 1})
\nonumber
\\
&+&
O\left((Tr^{\alpha})^2 n_0^{1-2\alpha}\right)  
+ O\left((Tr^{\alpha})^2(n_0 + n_1)^{1-2\alpha}\right) .\qquad
\label{eq-approx-q_i-case-1-other}
\end{eqnarray}
\end{lem}
%
%\begin{proof}
%The proof of this lemma is given in Appendix~\ref{appen-proof-lem-case-1}. 
%\end{proof}
%

\medskip

We next consider  case (ii) $1 \le \eta_{r,T}  \le n_1$, i.e.,
\begin{eqnarray}
\left(
1 + {1 \over n_0 + \eta_{r,T}}
\right)^{-\alpha}
\le
T
\left(
{ r
\over
n_0 + \eta_{r,T} + 1 }
\right)^{\alpha}
<
1.
\label{ineq-bar-i_T}
\end{eqnarray}
To proceed, we divide $q_{n_0, T}(r \mid n_1)$ into the following partial-sums:
\begin{eqnarray}
\overline{q}_{n_0, T}(r \mid n_1) &=& \sum_{m= \eta_{r,T}+1}^{n_1} \log\left(1 + T \left({r \over n_0 + m}\right)^{\alpha}\right),
\nonumber
\\
\underline{q}_{n_0, T}(r) &=& \sum_{m= 1}^{\eta_{r,T}} \log\left(1 + T \left({r \over n_0 + m}\right)^{\alpha}\right).
\label{eq-under-q_i(T)}
\end{eqnarray}
Lemma~\ref{lem-case-2} below shows  upper and lower bounds
for $\overline{q}_{n_0, T}(r \mid n_1)$. 
\begin{lem}\label{lem-case-2}
If $\eta_{r,T}$ given in 
(\ref{eq-def-over-i-T}) satisfies $1 \le \eta_{r,T} \le n_1$, 
there exist $\overline{q}_{\mathrm{lwr}}(\eta_{r,T} \mid n_1)$
and $\overline{q}_{\mathrm{upr}}(\eta_{r,T} \mid n_1)$ such that
\begin{equation}
\overline{q}_{\mathrm{lwr}}(\eta_{r,T} \mid n_1) < 
\overline{q}_{n_0, T}(r \mid n_1)
\le 
\overline{q}_{\mathrm{upr}}(\eta_{r,T} \mid n_1),
\label{eq-approx-xi_1}
\end{equation}
and
\begin{eqnarray}
\lefteqn{
\overline{q}_{\mathrm{upr}}(\eta_{r,T} \mid n_1) 
} &&
\nonumber
\\
&=& (\kappa_{1, \alpha} - \log 2 )
\left(n_0 + \eta_{r,T} + 1 \right) - \overline{\psi}_{n_0, n_1, T}(r) 
\nonumber
\\
&{}&-
{1\over 2}\log\left(1 + T\left({r \over n_0 + n_1}\right)^{\alpha}\right)
+ {1 \over 2} \log 2
\nonumber
\\
&{}&+
O\left((n_0 + \eta_{r,T})^{-1}\right)
+ O\left((n_0 + n_1)^{-1}\right),\qquad
\label{eq-upper-upper}
\end{eqnarray}
and
\begin{eqnarray}
\lefteqn{
\overline{q}_{\mathrm{lwr}}(\eta_{r,T} \mid n_1) 
= 
\overline{\varphi}(\eta_{r,T}) 
\left(
n_0 + \eta_{r,T} +1
\right)
- \overline{\psi}_{n_0, n_1, T}(r)
}\qquad \quad&&
\nonumber
\\
&-&
{1\over 2}\log\left(1 + T\left({r \over n_0 + n_1}\right)^{\alpha}\right)
\nonumber
\\
&+&
{1 \over 2}
\log\left(
1 + 
\left(
1 + {1 \over n_0 + \eta_{r,T}}
\right)^{-\alpha}
\right)
\nonumber
\\
&+&
O\left((n_0 + \eta_{r,T})^{-1}\right)
+ O\left((n_0 + n_1)^{-1}\right),\quad
\label{eq-upper-lower}
\end{eqnarray}
where $\kappa_{1, \alpha}$ is given in (\ref{eq-def-kappa}) and
\begin{eqnarray}
\overline{\varphi}(\eta_{r,T}) 
&=&
\sum_{k=1}^{\infty} {(-1)^{k+1} \over k(\alpha k - 1)}
\left(
1 + 
{1 \over n_0 + \eta_{r,T}}\right)^{-\alpha k },\quad
\label{eq-def-phi_0}
\\
\overline{\psi}_{n_0, n_1,T}(r) &= &
(n_0 + n_1)
 \sum_{k=1}^{\infty}
{(-1)^{k+1} T^k \over k (\alpha k - 1)}
\left(
{ r
\over
n_0 + n_1 
}
\right)^{\alpha k}.
\nonumber
\\
\label{eq-def-psi}
\end{eqnarray}
\end{lem}
%
%\begin{proof}
%The proof of this lemma is given in Appendix~\ref{appen-proof-lem-case-2}. 
%\end{proof}
%%

We next give upper and lower bounds for $\underline{q}_{n_0, T}(r)$. 
\begin{lem}\label{lem-case-3}
If $\eta_{r,T}$ given in (\ref{eq-def-over-i-T}) satisfies $1 \le \eta_{r,T} \le n_1$
and $\alpha \in \bbN$, then there exist $\underline{q}_{\mathrm{upr}}(\eta_{r,T})$
and 
$\underline{q}_{\mathrm{lwr}}(\eta_{r,T})$ such that
\begin{equation}
\underline{q}_{\mathrm{lwr}}(\eta_{r,T})
<
\underline{q}_{n_0, T}(r)
<
\underline{q}_{\mathrm{upr}}(\eta_{r,T}),
\label{eq-xi-2}
\end{equation}
and
\begin{eqnarray}
\lefteqn{
\underline{q}_{\mathrm{upr}}(\eta_{r,T}) 
= 
\left(\alpha 
+ \log 2 - \kappa_{2, \alpha}
\right) (n_0 + \eta_{r,T}) 
- \underline{\psi}_{n_0, T}(r) 
} &&
\nonumber
\\
&{}&
-
{1 \over 2}  \log \left(1 + {1 \over T}\left({n_0 \over r}\right)^{\alpha}\right)
+ \alpha \log\left(1 + {1 \over n_0 + \eta_{r,T}}\right)\eta_{r,T}
\nonumber
\\
&{}&
-
\alpha \left(n_0 + {1 \over 2}\right) \log(n_0 + \eta_{r,T})
+ \alpha \log {n_0! \over \sqrt{2 \pi}  }
\nonumber
\\
&{}&
+
{1 \over 2 }\log 2
+ O\left((n_0 + \eta_{r,T})^{-1}{1 \over T} \left({n_0 + \eta_{r,T}\over r} \right)^{\alpha}\right),
\label{eq-lower-upper}
\end{eqnarray}
and
\begin{eqnarray}
\lefteqn{
\underline{q}_{\mathrm{lwr}}(\eta_{r,T}) 
=
\left(
\alpha +
\underline{\varphi}(\eta_{r,T}) 
\right)(n_0 + \eta_{r,T})
- \underline{\psi}_{n_0, T}(r) 
} &&
\nonumber
\\
&{}&
-
{1 \over 2}  \log \left(1 + {1 \over T}\left({n_0 \over r}\right)^{\alpha}\right)
-  \alpha \left(n_0 + {1 \over 2}\right) \log(n_0 + \eta_{r,T})
\nonumber
\\
&{}&
+
\alpha \log {n_0 ! \over \sqrt{2\pi}}
+{1 \over 2}\log\left(
1 + \left(1 + {1 \over n_0 + \eta_{r,T}}\right)^{-\alpha}
\right)
\nonumber
\\
&{}&
+ O\left((n_0 + \eta_{r,T})^{-1}{1 \over T} \left({n_0 + \eta_{r,T}\over r} \right)^{\alpha}\right),
\label{eq-lower-lower}
\end{eqnarray}
where $\kappa_{2, \alpha}$ and $\underline{\psi}_{n_0, T}(r)$ are given in (\ref{eq-def-kappa}) 
and (\ref{eq-def-under-psi}), respectively and
\begin{eqnarray}
\underline{\varphi}(\eta_{r,T}) 
&=& 
\sum_{k=1}^{\infty} {(-1)^{k+1} \over k(\alpha k + 1)}
\left(
1 + 
{1 \over  n_0 + \eta_{r,T}}\right)^{-\alpha k },
\label{eq-def-phi_1}
%\\
%\underline{\psi}_{n_0, T}(r) &= &
%n_0 \sum_{k=1}^{\infty}
%{(-1)^{k+1}  \over k (\alpha k + 1) T^k}
%\left(
%{ n_0
%\over
%r
%}
%\right)^{\alpha k}.\quad
%\label{eq-def-under-psi}
\end{eqnarray}
\end{lem}
%
%\begin{proof}
%The proof of this lemma is given in Appendix~\ref{appen-proof-lem-case-3}. 
%\end{proof}
%%

Finally, we consider case (iii), i.e., the following holds, for any $m \in [1, n_1]$,
\begin{equation}
T\left({ r \over n_0 + m}\right)^{\alpha}\ge T\left({ r \over n_0 + n_1}\right)^{\alpha} > 1.
\label{eq-case-3-cond}
\end{equation}
\begin{lem}\label{lem-case-over}
If $\eta_{r,T}$ given in (\ref{eq-def-over-i-T}) satisfies $n_1 \le \eta_{r,T}$
and $\alpha \in \bbN$, then 
\begin{eqnarray}
\lefteqn{
q_{n_0, T}(r \mid n_1) 
=
\alpha n_1 \log T^{1 \over \alpha} r 
+ \alpha (n_0 + n_1) 
+ \alpha \log {n_0! \over \sqrt{2 \pi}  }
}  &&
\nonumber
\\
&{}&
- \alpha \left(n_0 + n_1 + {1 \over 2}\right) \log(n_0 + n_1)
\nonumber
\\
&{}&
+ {1 \over T} \left[ {n_0 + n_1 \over \alpha + 1} + {1 \over 2} \right] \left({n_0 + n_1 \over r}\right)^{\alpha}
\nonumber
\\
&{}&
-{1 \over T} \left[ {n_0  \over \alpha + 1} + {1 \over 2} \right] \left({n_0 \over r}\right)^{\alpha}
\nonumber
\\
&{}&
+
 O\left( (n_0 + n_1)^{-1}{1 \over T}\left({n_0 + n_1 \over r}\right)^\alpha \right)
\nonumber
\\
&{}&
+ O\left({1 \over T^2}\left({n_0 \over r}\right)^{2 \alpha}\right)
+ O\left(\log\left(1 + {1 \over n_0 + n_1}\right)\right).\qquad
\label{eq-case-over}
\end{eqnarray}
\end{lem}
%
%\begin{proof}
%The proof of this lemma is given in Appendix~\ref{proof-lem-case-over}. 
%\end{proof}
%%
%%
\medskip

We next derive approximate formulae for $q_{n_0, T}(r \mid n_0)$ 
on the basis of Lemmas~\ref{lem-case-1}--\ref{lem-case-over}. 
In cases (i) and (iii) where $\eta_{r,T}  = 0$ and $\eta_{r,T}  \ge n_1$,  
approximate formulae are directly obtained by applying 
Lemmas~\ref{lem-case-1} and \ref{lem-case-over}. 
Therefore, we only consider case (ii) $1 \le \eta_{r,T}  < n_1$
in what follows and explain how we derive the approximate formulae 
from Lemmas~\ref{lem-case-2} and \ref{lem-case-3}.

Note first that 
\begin{equation}
\log\left(1 + \left(1 + {1 \over n_0 + n_{r, T}}\right)^{-\alpha} \right) \approx \log 2,
\label{eq-approx-log2}
\end{equation}
for sufficiently large $n_0 + \eta_{r, T}$. 
Therefore, (\ref{eq-def-kappa}) and (\ref{eq-def-phi_0}) suggest that
\begin{eqnarray}
\overline{\varphi}(\eta_{r, T}) 
&=&
\sum_{k=1}^{\infty} (-1)^{k+1} 
\left[
{\alpha \over \alpha k - 1}
-
{1 \over k}
\right]
\left(
1 + 
{1 \over n_0 + \eta_{r,T}}\right)^{-\alpha k }
\nonumber
\\
&\approx& \kappa_{1, \alpha} - \log2 ,
\end{eqnarray}
for sufficiently large $n_0 + \eta_{r, T}$. In addition, $\psi_{n_0, n_1, T}(r)$ can be
rewritten as [see (\ref{eq-def-psi})]
\begin{eqnarray}
\overline{\psi}_{n_0, n_1, T}(r)
&=& {(n_0 + n_1) T\over \alpha -1}\left({r \over n_0 + n_1}\right)^{\alpha}
\nonumber
\\
&+& O\left( (n_0 + n_1)T^2 \left({r \over n_0 + n_1}\right)^{2 \alpha}
\right).
\label{eq-beta-approx-1}
\nonumber
\end{eqnarray}
Furthermore, from the definition [see (\ref{eq-def-over-i-T})],  $\eta_{r, T}$ can be approximated as
\begin{equation}
\eta_{r, T} \approx T^{1/\alpha} r - n_0.
\label{eq-app-eta}
\end{equation}
Therefore, by combining these facts with (\ref{eq-approx-xi_1})--(\ref{eq-def-psi}), 
we can obtain the following approximation for $\overline{q}_{n_0, T}(r \mid n_1)$:
\begin{eqnarray}
\lefteqn{
\overline{q}_{n_0, T}(r \mid n_1) 
\approx
(\kappa_{1, \alpha} - \log 2) T^{1 \over \alpha}r 
+ \kappa_{1, \alpha} + {1 \over 2} \log 2
} &&
\nonumber
\\
&-&
{T r^{\alpha} \over (\alpha -1)(n_0 + n_1)^{\alpha -1}}
+
{1 \over 2} \log\left(1 + T\left({r \over n_0 + n_1}\right)^{\alpha}\right)
. 
\nonumber
\\
\label{eq-approx-over-q_r(T)}
\end{eqnarray}
Similar to the above arguments, it can be said that [see (\ref{eq-def-kappa})]
\begin{eqnarray}
\underline{\varphi}(\eta_{r,T})  
&=&
\sum_{k=1}^{\infty} (-1)^{k+1} 
\left[
{1 \over k}
-
{\alpha \over \alpha k + 1}
\right]
\left(
1 + 
{1 \over n_0 + \eta_{r,T}}\right)^{-\alpha k }
\nonumber
\\
&\approx&
\log 2 - 
\kappa_{2, \alpha}, 
\nonumber
\end{eqnarray}
for sufficiently large $n_0 + \eta_{r,T}$. 
Applying this, (\ref{eq-approx-log2}) and (\ref{eq-app-eta})
to (\ref{eq-xi-2})--(\ref{eq-def-phi_1}), we can 
approximate $\underline{q}_{n_0, T}(r)$ as
\begin{eqnarray}
\underline{q}_{n_0, T}(r) &\approx&
(\alpha + \log 2 - \kappa_{2, \alpha}) T^{1 \over \alpha} r
- \underline{\psi}_{n_0, T}(r) 
\nonumber
\\
&-&
{1 \over 2}  \log \left(1 + {1 \over T}\left({n_0 \over r}\right)^{\alpha}\right)
- \alpha \left(n_0 + {1 \over 2}\right) \log T^{1 \over \alpha} r
\nonumber
\\
&+& 
\alpha \log {n_0 ! \over \sqrt{2 \pi}}
+ {1 \over 2}\log 2.
\label{eq-approx-under-q_r(T)}
\end{eqnarray}
As a result, by combining (\ref{eq-approx-over-q_r(T)}) and 
(\ref{eq-approx-under-q_r(T)}), 
we obtain (\ref{approx-fand}). 
%the following approximate formulae for $q_{n_0, T}(r)$. 
%

\section{Proofs}\label{appen-proof}
\subsection{Proof of Lemma~\ref{lem-case-1}}\label{appen-proof-lem-case-1}
It follows from  Taylor's theorem that
\begin{eqnarray}
\log \left( 1 + x^{\alpha}\right)
&=& 
\sum_{k=1}^\infty{ (-1)^{k+1} \over k} ( x^{\alpha})^k,
\qquad  x^{\alpha} < 1,
\label{eq-appr-log-x<1}
\end{eqnarray}
from which and (\ref{eq-def-q_i(T)}) we obtain
\begin{eqnarray}
q_{n_0, T}(r \mid n_1)
&=&
\sum_{m= n_0+1}^{n_0 + n_1} \sum_{k=1}^\infty {(-1)^{k+1} T^k \over k }
\left({r \over m}\right)^{\alpha k }
\nonumber
\\
&=&
\sum_{k=1}^\infty {(-1)^{k+1} T^k r^{\alpha k} \over k }
\sum_{m= n_0+1}^{n_0 + n_1}  \left({1 \over m}\right)^{\alpha k }.\qquad
\label{eq-sum-bar-i-N-log-1}
\end{eqnarray}
%}
%
Furthermore, 
applying the Euler-Maclaurin summation formula
to the Riemann zeta function leads to (see e.g., Section~6.4 in \cite{Edwa01} for details)
\begin{eqnarray}
\sum_{m=1}^n {1 \over m^{\alpha}} = \zeta(\alpha) 
+ {n^{1-\alpha} \over 1 - \alpha} - {n^{-\alpha} \over 2}
  + O(n^{-\alpha - 1}),
\label{eq-approx-zeta}
\end{eqnarray}
where $\zeta(\alpha)$ is given in (\ref{eq-def-zeta}). 
Therefore, applying (\ref{eq-approx-zeta}) to (\ref{eq-sum-bar-i-N-log-1})
and letting $n_0 = 0$ yields 
\begin{eqnarray}
q_{0, T}(r \mid n_1) = \sum_{k=1}^{\infty}{(-1)^{k+1} (T r^{\alpha})^k \over k }
\left[\zeta(\alpha k) + O(n_1^{1 - \alpha k})
\right],
\nonumber
\end{eqnarray}
from which, we obtain (\ref{eq-approx-q_i-case-1}). 
On the other hand, if $n_0 \ge 1$, by using (\ref{eq-approx-zeta}),
the last summation in (\ref{eq-sum-bar-i-N-log-1}) can be rewritten as 
\begin{eqnarray}
\lefteqn{
\sum_{m= n_0+1}^{n_0 + n_1}  \left({1 \over m}\right)^{\alpha k }
= 
\sum_{m= 1}^{n_0 + n_1}  \left({1 \over m}\right)^{\alpha k }
- 
\sum_{m= 1}^{n_0 }  \left({1 \over m}\right)^{\alpha k }
} && 
\nonumber
\\
&=&
{n_0^{1-\alpha k } 
- (n_0 + n_1)^{1-\alpha k }
 \over \alpha k -1}
+
{n_0^{-\alpha k }  - (n_0 + n_1)^{-\alpha k } \over 2}
\nonumber
\\
&{}&
+ O(n_0^{-\alpha k - 1})
+ O((n_0 + n_1)^{-\alpha k - 1}).
\nonumber
\end{eqnarray}
Thus, plugging the above into (\ref{eq-sum-bar-i-N-log-1}), 
we obtain
\begin{eqnarray}
\lefteqn{
q_{n_0, T}(r \mid n_1)
=
\sum_{k=1}^\infty {(-1)^{k+1} (T r^{\alpha })^k \over k }
} &&
\nonumber
\\
&\times&
\left[
{n_0^{1-\alpha k } 
- (n_0 + n_1)^{1-\alpha k }
 \over \alpha k -1}
+
{n_0^{-\alpha k }  - (n_0 + n_1)^{-\alpha k } \over 2}
\right.
\nonumber
\\
&+&
\left.
O(n_0^{-\alpha k-1})
+ O((n_0 + n_1)^{-\alpha k-1})\right],
\nonumber
\end{eqnarray}
which leads to (\ref{eq-approx-q_i-case-1-other}). 
\subsection{Proof of Lemma~\ref{lem-case-2}}\label{appen-proof-lem-case-2}
Note first that $T({r \over m})^{\alpha} < 1$
for any $m > n_0 + \eta_{r,T}$ according to the definition of $\eta_{r,T}$ [see (\ref{ineq-bar-i_T})]. 
Thus, similar to the derivation of (\ref{eq-sum-bar-i-N-log-1}), 
we obtain 
\begin{equation}
\overline{q}_{n_0, T}(r \mid n_1) =
\sum_{k=1}^\infty {(-1)^{k+1} T^k r^{\alpha k} \over k }
\sum_{m= n_0 + \eta_{r,T}+1}^{n_0 + n_1}  \left({1 \over  m}\right)^{\alpha k }.\quad
\label{eq-upper-q-add-1}
\end{equation}
Furthermore, it follows from (\ref{eq-approx-zeta}) that, for any $k \in \bbN$,
\begin{eqnarray}
\lefteqn{
\sum_{m= n_0 + \eta_{r, T} + 1}^{n_0 + n_1} \left({1 \over m}\right)^{\alpha k }
=
\sum_{m= 1}^{n_0 + n_1}  \left({1 \over m}\right)^{\alpha k }
-
\sum_{m= 1}^{n_0 + \eta_{r, T}} \left({1 \over m}\right)^{\alpha k }
}&&
\nonumber
\\
&=&
 {(n_0 + \eta_{r,T} + 1)^{1 - \alpha k} \over  \alpha k- 1} - {(n_0 + n_1)^{1 - \alpha k} \over  \alpha k- 1}
\nonumber
\\ 
&{}&
+ {(n_0 + \eta_{r,T} + 1)^{ - \alpha k} \over  2} - {(n_0 + n_1)^{ - \alpha k} \over  2}
\nonumber
\\
&{}&
+ O((n_0 + \eta_{r, T}+1)^{-\alpha k-1}) + O((n_0 + n_1)^{-\alpha k-1}).\qquad
\label{eq-xi-add-1}
\end{eqnarray}
Thus, applying (\ref{eq-appr-log-x<1}) and (\ref{eq-xi-add-1})
to (\ref{eq-upper-q-add-1}) leads to 
\begin{eqnarray}
\lefteqn{
\overline{q}_{n_0, T}(r \mid n_1)
= 
\sum_{k=1}^\infty 
{(-1)^{k+1}  \over k }
}\qquad && 
\nonumber
\\
&{}&
\times
\left[\left({n_0 + \eta_{r,T} + 1   \over \alpha k - 1} + {1\over 2}\right)
T^k \left({r \over n_0 + \eta_{r,T} + 1}\right)^{\alpha k }
\right.
\nonumber
\\
&{}&-
\left({n_0 + n_1 \over \alpha k - 1} + {1 \over 2} \right)
T^k\left({r \over n_0 + n_1}\right)^{\alpha k }
\nonumber
\\
&{}&+
O\left((n_0 + \eta_{r, T}+1)^{-1}T^k\left({r \over n_0 + \eta_{r,T} + 1}\right)^{\alpha k }\right)
\nonumber
\\
&{}&
\left.
+
O\left((n_0 + n_1)^{-1}T^k\left({r \over n_0 + n_1}\right)^{\alpha k }\right) 
\right].
\label{eqn-up-barq-1}
\end{eqnarray}
In addition, (\ref{ineq-bar-i_T}) suggests that
\begin{eqnarray}
\lefteqn{
\sum_{k=1}^\infty 
{(-1)^{k+1} T^k  \over k }
\left({n_0 + \eta_{r,T} + 1   \over \alpha k - 1} + {1\over 2}\right)
\left({r \over n_0 + \eta_{r,T} + 1}\right)^{\alpha k }
}\qquad&&
\nonumber
\\
&<&
\sum_{k=1}^\infty 
{(-1)^{k+1} T^k  \over k }
\left[
{n_0 + \eta_{r,T} + 1   \over \alpha k - 1} + {1\over 2}
\right]
\nonumber
\\
&=&
\sum_{k=1}^\infty 
{(-1)^{k+1} \over k (\alpha k - 1) }
(n_0 + \eta_{r,T} + 1)
+ {1\over 2}\log2
\nonumber
\\
&=&
\sum_{k=1}^\infty 
(-1)^{k+1}
\left[
{\alpha  \over \alpha k - 1}
-
{1 \over k}
\right]
(n_0 + \eta_{r,T} + 1)
+ {1\over 2}\log2
\nonumber
\\
&=& 
(\kappa_{1, \alpha} - \log 2)
(n_0 + \eta_{r,T} + 1)
+ {1\over 2}\log2,
\qquad\quad
\label{eq-add-barq-up-tmp}
\end{eqnarray}
where we use $\sum_{k=1}^{\infty}(-1)^{k+1}/k = \log2$ in the first and last equalities
and $\kappa_{1, \alpha}$ is given in (\ref{eq-def-kappa}). 
Furthermore, using (\ref{eq-appr-log-x<1}), we  obtain
\begin{eqnarray}
\lefteqn{
\sum_{k=1}^{\infty} {(-1)^{k+1} T^k \over k} 
\left({r \over n_0 + n_1}\right)^{\alpha k }
}\qquad \quad&&
\nonumber
\\
&=&
\log\left(1 + T\left({r \over n_0 + n_1}\right)^{\alpha}\right).
\label{eq-sum-r/n_0+n_1}
\end{eqnarray}
Therefore, combining (\ref{eq-def-psi}), (\ref{eq-add-barq-up-tmp}), and (\ref{eq-sum-r/n_0+n_1})
with (\ref{eqn-up-barq-1}), 
the second inequality in (\ref{eq-approx-xi_1}) and (\ref{eq-upper-upper}),
i.e., the upper bound for $\overline{q}_{n_0, T}(r \mid n_1)$, is proved.

We next prove the first inequality in (\ref{eq-approx-xi_1}), i.e., the lower bound. 
Similar to the derivation of (\ref{eq-add-barq-up-tmp}), 
combining (\ref{ineq-bar-i_T}) with (\ref{eq-xi-add-1}) yields
\begin{eqnarray}
\lefteqn{
\sum_{k=1}^\infty 
{(-1)^{k+1} T^k  \over k }
\left({n_0 + \eta_{r,T} + 1   \over \alpha k - 1} + {1\over 2}\right)
\left({r \over n_0 + \eta_{r,T} + 1}\right)^{\alpha k }
}&&
\nonumber
\\
&\ge&
\sum_{k=1}^\infty 
{(-1)^{k+1} \over k }
\left(
{n_0 + \eta_{r,T} + 1 \over \alpha k - 1} + {1 \over 2}
\right)
\left(
1 + {1 \over n_0 + \eta_{r,T}}
\right)^{-\alpha k}
\nonumber
\\
&=&
\overline{\varphi}(\eta_{r,T}) 
\left(
n_0 + \eta_{r,T} +1
\right)
\nonumber
\\
&{}&
+ 
{1 \over 2}
\log\left(
1 + 
\left(
1 + {1 \over n_0 + \eta_{r,T}}
\right)^{-\alpha}
\right),
\nonumber
\end{eqnarray}
where we use (\ref{eq-def-phi_0}) and (\ref{eq-appr-log-x<1}) in the equality. 
Consequently, substituting this, (\ref{eq-def-psi}), and (\ref{eq-sum-r/n_0+n_1})
into (\ref{eqn-up-barq-1}) leads to (\ref{eq-approx-xi_1}) and (\ref{eq-upper-lower}).

\subsection{Proof of Lemma~\ref{lem-case-3}}\label{appen-proof-lem-case-3}
From Taylor's theorem, we obtain for $x > 1$, 
\begin{equation}
\log \left( 1 +  x^{\alpha}\right)
= 
\log  x^{\alpha} +
\sum_{k=1}^\infty {(-1)^{k+1} \over k} {1 \over  x^{\alpha k}}.
\nonumber
\end{equation}
Since $T ({r \over n_0 + m})^{\alpha} > 1$ for any $m \in [1, \eta_{r,T}]$,
applying the above equation to (\ref{eq-under-q_i(T)}) leads to 
\begin{eqnarray}
\lefteqn{
\underline{q}_{n_0, T}(r)
} &
\nonumber
\\
&=&
\sum_{m=n_0 + 1}^{n_0 + \eta_{r,T}}\log T \left({r \over m}\right)^{\alpha}
+ \sum_{m=n_0 + 1}^{n_0 + \eta_{r,T}} \sum_{k=1}^\infty {(-1)^{k+1}  \over k T^k} \left({m \over r}\right)^{\alpha k }
\nonumber
\\
&=&
\eta_{r,T} \log T + \alpha \left(\eta_{r,T} \log r - \log (n_0 + \eta_{r,T})_{\eta_{r,T}}  \right)
\nonumber
\\
&{}& + 
\sum_{k=1}^\infty {(-1)^{k+1} \over k T^k} \left({1 \over r}\right)^{\alpha k }
\sum_{m=1}^{\eta_{r,T}}(n_0 + m)^{\alpha k },
\label{eq-sum-log-m<i-1}
\end{eqnarray}
where $(x)_k$ $(k \in \bbN)$ denotes the falling sequential product such that
\[
(x)_k = x (x - 1) \cdots (x - k +1). 
\]
It follows from Faulhaber's formula (see e.g., \cite{Conw96}) that for any $n \in \bbN$,
\begin{eqnarray}
\lefteqn{
\sum_{m=1}^n (n_0 + m)^{\alpha k}
} &&
\nonumber
\\
&=& 
{1 \over \alpha k+ 1}
\sum_{j=0}^{\alpha k}{\alpha k+1 \choose j} B_j 
 \left[(n_0 + n)^{\alpha k+1-j} - (n_0)^{\alpha k+1-j}\right]
\nonumber
\\
&=&
{1 \over \alpha k+ 1}
\sum_{j=0}^{\alpha k}{\alpha k+1 \choose j} B_j 
\nonumber
\\
&{}&
\times
(n_0 + n)^{\alpha k+1-j} \left[1 - \left(n_0 \over n_0 + n\right )^{\alpha k+1-j}\right],
\label{eq-Faulhaber}
\end{eqnarray}
where $B_j$'s are the Bernoulli numbers  such that
\[
B_0 = 1,\qquad
B_j = \sum_{k=0}^{j-1} (-1)^k {j +1 \choose k}B_k,
\quad j \ge 1.
\]
Substituting (\ref{eq-Faulhaber}) into the second term on the 
right-hand side of (\ref{eq-sum-log-m<i-1}) yields
\begin{eqnarray}
\lefteqn{
\sum_{k=1}^\infty {(-1)^{k+1} \over k T^k} \left({1 \over r}\right)^{\alpha k }
\sum_{m=1}^{\eta_{r,T}}(n_0 + m)^{\alpha k}
} && 
\nonumber
\\
&=&
\sum_{k=1}^\infty {(-1)^{k+1} \over k T^k} \left({1 \over r}\right)^{\alpha k }
{1 \over \alpha k + 1} \sum_{j=0}^{\alpha k}{\alpha k +1 \choose j} B_j
\nonumber
\\
&{}&
\times
(n_0 + \eta_{r,T})^{\alpha k+1-j} \left[1 - \left(n_0 \over n_0 + \eta_{r,T}\right )^{\alpha k+1-j}\right]
\nonumber
\\
&=& 
\sum_{k=1}^\infty {(-1)^{k+1} \over k T^k}
\left[
\left(
{n_0 + \eta_{r,T} \over \alpha k + 1}
+ {1 \over 2}
\right)
\left({n_0 + \eta_{r,T} \over r}\right)^{\alpha k }
\right.
\nonumber
\\
&{}&
-
\left(
{n_0  \over \alpha k + 1}
+ {1 \over 2}
\right)
\left({n_0 \over r}\right)^{\alpha k }
+
{r^{- \alpha k} \over \alpha k + 1} \sum_{j=2}^{\alpha k}{\alpha k +1 \choose j} B_j
\nonumber
\\
&{}&
\times
\left.
(n_0 + \eta_{r,T})^{\alpha k+1-j} \left[1 - \left(n_0 \over n_0 + \eta_{r,T}\right )^{\alpha k+1-j}\right]
\right].
\nonumber
\\
\label{eq-sum-Bj-Qi}
\end{eqnarray}
Note that (\ref{eq-def-over-i-T}) and $1 \le \eta_{r,T}$  suggest that
\begin{eqnarray}
\left(
1 + {1 \over n_0 + \eta_{r,T}}
\right)^{-\alpha}
<
{1 \over T}
\left(
{n_0 + \eta_{r,T}
\over
r }
\right)^{\alpha}
\le
1.
\label{ineq-underbar-i_T}
\end{eqnarray}
Note also that (see (\ref{eq-def-under-psi}) and (\ref{eq-appr-log-x<1}))
\begin{eqnarray}
\lefteqn{
\sum_{k=1}^\infty {(-1)^{k+1} \over k T^k}
\left(
{n_0  \over \alpha k + 1}
+ {1 \over 2}
\right)
\left({n_0 \over r}\right)^{\alpha k }
}\qquad &&
\nonumber
\\
&=&
\underline{\psi}_{n_0, T}(r) 
+
{1 \over 2}  \log \left(1 + {1 \over T}\left({n_0 \over r}\right)^{\alpha}\right).
\label{eq-add-underbar-psi-tmp}
\end{eqnarray}
Thus, by using (\ref{ineq-underbar-i_T}) and (\ref{eq-add-underbar-psi-tmp}), we obtain
\begin{eqnarray}
\lefteqn{
\sum_{k=1}^\infty {(-1)^{k+1} \over k T^k}
\left[
\left(
{n_0 + \eta_{r,T} \over \alpha k + 1}
+ {1 \over 2}
\right)
\left({n_0 + \eta_{r,T} \over r}\right)^{\alpha k }
\right.} &&
\nonumber
\\
&{}&
\left.
-
\left(
{n_0  \over \alpha k + 1}
+ {1 \over 2}
\right)
\left({n_0 \over r}\right)^{\alpha k }
\right]
\nonumber
\\
&\le&
\sum_{k=1}^\infty {(-1)^{k+1} \over k }\left[
{n_0 + \eta_{r,T} \over \alpha k + 1}
+ {1 \over 2}
\right]
- 
\underline{\psi}_{n_0, T}(r) 
\nonumber
\\
&{}&
-
{1 \over 2}  \log \left(1 + {1 \over T}\left({n_0 \over r}\right)^{\alpha}\right)
\nonumber
\\
&=&
\left(
\log2 - \kappa_{2, \alpha}
\right)
(n_0 + \eta_{r,T})
+ {1 \over 2} \log 2
- 
\underline{\psi}_{n_0, T}(r) 
\nonumber
\\
&{}&
-
{1 \over 2}  \log \left(1 + {1 \over T}\left({n_0 \over r}\right)^{\alpha}\right),
\label{eq-lower-upper-tmp}
\end{eqnarray}
where $\kappa_{2,\alpha}$ is given in (\ref{eq-def-kappa})
and we use $\sum_{k=1}^{\infty}(-1)^{k+1}/k = \log2$ in the equality. 
Similarly, we have
\begin{eqnarray}
\lefteqn{
\sum_{k=1}^\infty {(-1)^{k+1} \over k T^k}
\left[
\left(
{n_0 + \eta_{r,T} \over \alpha k + 1}
+ {1 \over 2}
\right)
\left({n_0 + \eta_{r,T} \over r}\right)^{\alpha k }
\right.} &&
\nonumber
\\
&{}&
\left.
-
\left(
{n_0  \over \alpha k + 1}
+ {1 \over 2}
\right)
\left({n_0 \over r}\right)^{\alpha k }
\right]
\nonumber
\\
&\ge&
\sum_{k=1}^\infty {(-1)^{k+1} \over k }\left[
{n_0 + \eta_{r,T} \over \alpha k + 1}
+ {1 \over 2}
\right]
\left(
1 + {1 \over  n_0 + \eta_{r,T}}
\right)^{- \alpha k}
\nonumber
\\
&{}&
-
\underline{\psi}_{n_0, T}(r) 
-
{1 \over 2}  \log \left(1 + {1 \over T}\left({n_0 \over r}\right)^{\alpha}\right)
\nonumber
\\
&=&
\underline{\varphi}(\eta_{r,T}) 
(n_0 + \eta_{r,T})
+
{1 \over 2}
\log\left(
1 + \left(1 + {1 \over n_0 + \eta_{r,T}}\right)^{-\alpha}
\right)
\nonumber
\\
&{}&
- \underline{\psi}_{n_0, T}(r) 
-
{1 \over 2}  \log \left(1 + {1 \over T}\left({n_0 \over r}\right)^{\alpha}\right)
,
\label{eq-lower-lower-tmp}
\end{eqnarray}
where the equality follows from  (\ref{eq-def-phi_1}) and (\ref{eq-add-underbar-psi-tmp}). 
Furthermore, it follows from (\ref{ineq-underbar-i_T}) that
\begin{eqnarray}
\lefteqn{
\sum_{k=1}^\infty {(-1)^{k+1} \over T^k k (\alpha k + 1)}
\sum_{j=2}^{\alpha k}{\alpha k +1 \choose j} B_j
} &&
\nonumber
\\
&\times&
{(n_0 + \eta_{r,T})^{\alpha k+1-j} \over r^{\alpha k}}\left[1 - \left(n_0 \over n_0 + \eta_{r,T}\right )^{\alpha k+1-j}\right]
\nonumber
\\
&=&O\left((n_0 + \eta_{r, T})^{-1} {1 \over T}\left({n_0 + \eta_{r,T} \over r}\right)^{\alpha}\right).
\label{ineq-B-upper}
\end{eqnarray}
Therefore, the lower and upper bounds for the last term on the right-hand
side of (\ref{eq-sum-log-m<i-1}) are shown. 

We next consider the first and second terms in (\ref{eq-sum-log-m<i-1}). 
It follows from Stirling's formula that
\begin{eqnarray}
\lefteqn{
\eta_{r,T}  \log r  - \log (n_0 + \eta_{r,T})_{\eta_{r,T}}
} &&
\nonumber
\\
&=&
\eta_{r,T}  \log r  - \log \sqrt{2 \pi (n_0 + \eta_{r,T})} 
\nonumber
\\
&{}&
-
(n_0 + \eta_{r,T}) \log\left({n_0 + \eta_{r,T} \over e}\right)
+ \log n_0 !
\nonumber
\\
&{}&
+ O\left(\log\left(1 + {1 \over n_0 + \eta_{r,T}}\right)\right)
\nonumber
\\
&=&
n_0 + \eta_{r,T} + 
\eta_{r,T} \log\left({r \over  n_0 + \eta_{r,T}}\right)
+\log {n_0! \over \sqrt{2 \pi}  }
\nonumber
\\
&-&
\left(n_0 + {1 \over 2}\right) \log(n_0 + \eta_{i,T})
+ O\left(\log\left(1 + {1 \over n_0 + \eta_{r,T}}\right)\right).
\nonumber
\\
\label{eq-stirling}
\end{eqnarray}
Note that (\ref{ineq-underbar-i_T}) suggests that
\begin{eqnarray}
\log\left({r \over  n_0 + \eta_{r,T}}\right)
&>&
\log
{1 \over T^{1 /\alpha}},
\label{ineq-log(r/n0+eta)-low}
\\
\log\left({r \over  n_0 + \eta_{r,T}}\right)
&\le&
\log
{1 \over T^{1 /\alpha}}
\left(
1 + {1 \over n_0 + \eta_{r,T}}
\right).\quad
\label{ineq-log(r/n0+eta)-up}
\end{eqnarray}
As a result, combining (\ref{ineq-log(r/n0+eta)-up}) with (\ref{eq-stirling})
and using this, (\ref{eq-lower-upper-tmp}), (\ref{ineq-B-upper}), and (\ref{eq-sum-Bj-Qi}), 
we obtain (\ref{eq-xi-2}) and (\ref{eq-lower-upper}). 
Similarly, 
by combining (\ref{ineq-log(r/n0+eta)-low}) with (\ref{eq-stirling}) 
and applying this, (\ref{eq-lower-lower-tmp}), and (\ref{ineq-B-upper})
into (\ref{eq-sum-Bj-Qi}), we obtain (\ref{eq-xi-2}) and  (\ref{eq-lower-lower}).

\subsection{Proof of Lemma~\ref{lem-case-over}}\label{proof-lem-case-over}

Similar to the derivation of (\ref{eq-sum-log-m<i-1}), it follows from (\ref{eq-case-3-cond}) that
\begin{eqnarray}
\lefteqn{
q_{n_0, T}(r) 
} &&
\nonumber
\\
&=&
\sum_{m=n_0 + 1}^{n_0 + n_1}\log T \left({r \over m}\right)^{\alpha}
+ \sum_{m=n_0 + 1}^{n_0 + n_1} \sum_{k=1}^\infty {(-1)^{k+1}  \over k T^k} \left({m \over r}\right)^{\alpha k }
\nonumber
\\
&=&
n_1 \log T + \alpha \left(n_1 \log r - \log (n_0 + n_1)_{n_1}  \right)
\nonumber
\\
&{}& + 
\sum_{k=1}^\infty {(-1)^{k+1} \over k T^k} \left({1 \over r}\right)^{\alpha k }
\sum_{m=1}^{n_1 }(n_0 + m)^{\alpha k }.
\label{eq-over-tmp-1}
\end{eqnarray}
%
%}
Applying the same technique in the derivation of (\ref{eq-sum-Bj-Qi}) to 
the second term in (\ref{eq-over-tmp-1}) yields
\begin{eqnarray}
\lefteqn{
\sum_{k=1}^\infty {(-1)^{k+1} \over k T^k} \left({1 \over r}\right)^{\alpha k }
\sum_{m=1}^{n_1}(n_0 + m)^{\alpha k}
} && 
\nonumber
\\
&=& 
\sum_{k=1}^\infty {(-1)^{k+1} \over k  T^k}
\left[
\left(
{n_0 + n_1 \over \alpha k + 1}
+ {1 \over 2}
\right) \left({n_0 + n_1 \over r}\right)^{\alpha k }
\right.
\nonumber
\\
&{}&
-
\left(
{n_0  \over \alpha k + 1}
+ {1 \over 2}
\right) \left({n_0  \over r}\right)^{\alpha k }
\nonumber
\\
&{}&
\left.
+ O(r^{-\alpha k}n_0^{\alpha k-1})
+ O(r^{-\alpha k}(n_0 + n_1)^{\alpha k-1})
\right]
\nonumber
\\
&=& 
{1 \over T} \left[ {n_0 + n_1 \over \alpha + 1} + {1 \over 2} \right] \left({n_0 + n_1 \over r}\right)^{\alpha}
\nonumber
\\
&{}&
-
{1 \over T} \left[ {n_0  \over \alpha + 1} + {1 \over 2} \right] \left({n_0 \over r}\right)^{\alpha}
+
O\left({1 \over T}{n_0^{\alpha -1} \over r^{\alpha}}\right)
\nonumber
\\
&{}&
+ O\left((n_0 + n_1)^{-1}{1 \over T}\left({n_0 + n_1 \over r}\right)^{\alpha}\right).
\label{eq-over-sum-second-add}
\end{eqnarray}
In addition, similar to (\ref{eq-stirling}), Stirling's formula leads to 
\begin{eqnarray}
\lefteqn{
n_1  \log r  - \log (n_0 + n_1)_{n_1}
} &&
\nonumber
\\
&=&
n_0 + n_1 + 
n_1 \log\left({r \over  n_0 + n_1}\right)
+\log {n_0! \over \sqrt{2 \pi}  }
\nonumber
\\
&{}&
- \left(n_0 + {1 \over 2}\right) \log(n_0 + n_1)
+ O\left(\log\left(1 + {1 \over n_0 + n_1}\right)\right).
\nonumber
\end{eqnarray}
Substituting this and (\ref{eq-over-sum-second-add}) into (\ref{eq-over-tmp-1}) results in (\ref{eq-case-over}).

%\section*{Acknowledgment}

%\ifCLASSOPTIONcaptionsoff
%  \newpage
%\fi



%\bibliographystyle{IEEEtran}

\end{document}